\documentclass[twocolumn,groupedaddress,aps,pra,groupedaddress,
amsmath,amssymb,showpacs]{revtex4-1}
\usepackage[colorlinks,linkcolor=blue,anchorcolor=blue,
         	citecolor=blue,bookmarksnumbered]{hyperref}
\usepackage{graphicx,epsfig,float,epstopdf}
\usepackage{amssymb,amsmath,subeqnarray,mathrsfs,
           amsfonts,bm,makecell,cases,color,extpfeil,lipsum}
\usepackage{dcolumn,comment}\allowdisplaybreaks
\usepackage{tikz}
\usepackage{etoolbox}
\newcommand{\circled}[2][]{\tikz[baseline=(char.base)]
    {\node[shape = circle, draw, inner sep = 0.1pt]
    (char) {\phantom{\ifblank{#1}{#2}{#1}}};%
    \node at (char.center) {\makebox[0pt][c]{#2}};}}
\robustify{\circled} 
\usepackage{ulem}

\begin{document}
\newcommand{\N}{\mathcal{N}}
\newcommand{\MHz}{2\pi\times\,\rm{MHz}}
\newcommand{\red}[1]{\textcolor{red}{{}#1}}
\newcommand{\blue}[1]{\textcolor{blue}{{}#1}}
\hoffset=-10pt

\title{Structural Phase Transitions of Optical Patterns in Atomic Gases with Microwave Controlled Rydberg Interactions}
\author{Zeyun Shi$^{1}$, Weibin Li$^{2,4}$, and Guoxiang Huang$^{1,3,5}$
}
\affiliation{
$^1$State Key Laboratory of Precision Spectroscopy, East China Normal University, Shanghai 200062, China\\
$^2$School of Physics and Astronomy, University of Nottingham, Nottingham, NG7 2RD, UK\\
$^3$NYU-ECNU Joint Institute of Physics, New York University Shanghai, Shanghai 200062, China\\
$^4$Centre for the Mathematics and Theoretical Physics of Quantum Non-equilibrium Systems, University of Nottingham, Nottingham, NG7 2RD, UK\\
$^5$Collaborative Innovation Center of Extreme Optics, Shanxi University, Taiyuan, Shanxi 030006, China
}
\date{\today}

\begin{abstract}

Spontaneous symmetry breaking and formation of self-organized  structures in nonlinear systems are intriguing and important phenomena in nature. Advancing such research to new nonlinear optical regimes is of much interest for both fundamental physics and practical applications. Here we propose a scheme to realize optical pattern formation in a cold Rydberg atomic gas via electromagnetically induced transparency. We show that, by coupling two Rydberg states with a microwave field (microwave dressing), the nonlocal Kerr nonlinearity of the Rydberg gas can be enhanced significantly and may be tuned actively. Based on such nonlocal Kerr nonlinearity, we demonstrate that a plane-wave state of probe laser field can undergo a modulation instability (MI) and hence spontaneous symmetry breaking, which
may result in the emergence of various self-organized optical patterns. Especially, we find that a hexagonal lattice pattern (which is the only optical pattern when the microwave dressing is absent) may develop into several types of square lattice ones when the microwave dressing is applied; moreover, as a outcome of the MI the formation of nonlocal optical solitons is also possible in the system. Different from  earlier studies, the optical patterns and nonlocal optical solitons found here can be flexibly manipulated by adjusting the effective probe-field intensity, nonlocality degree of the Kerr nonlinearity, and the strength of the microwave field. Our work opens a route for versatile controls of self-organizations and structural phase transitions of laser light,
which may have potential applications in optical information processing and transmission.

\pacs{42.65.Sf, 42.65.Tg, 32.80.Ee, 42.50.Gy}

\end{abstract}

\maketitle

\section{Introduction}\label{sec1}

Symmetry breaking and formation of ordered structures~(patterns) in spatially extended dissipative systems driven away from equilibrium via some instability mechanisms are very interesting and important phenomena, occurring widely in physics, chemistry, biology, cosmology, and even economics and sociology,
etc.~\cite{Nicolis1977,Haken1987,Murray1989,Cross1993,Bowman1998,
Weidman2000,Cross2009}.  Well-known instability mechanisms for pattern formations include the Rayleigh-B\'{e}nard instability in thermal fluid convection~\cite{Benard1901,Drazin1981}, the
Taylor-Couette instability in rotating fluids~\cite{Taylor1923}, the
electrohydrodynamic instability in nematic liquid crystals~\cite{Dubois1978},
and the Faraday instability for parametric waves~\cite{Faraday1831}; other typical examples are the lasing instability in laser devices~\cite{Newell1990,Arecchi1999,Lugiato1999}, the
Mullins-Sekerka instability for solidification pattern growth (e.g., snowflakes)~\cite{Mullins1964},
and the Turing instability for structures created in chemical reaction and living systems (e.g., animal coats)~\cite{Turing1952}. For details, see Refs.~\cite{Nicolis1977,Haken1987,Murray1989,Cross1993,Bowman1998,
Weidman2000,Cross2009} and references cited therein.
One of main characters of these pattern forming systems is that an external drive (stress) must be applied, and the induced instability trigers symmetry-breaking causing the dissipative structures appearing immediately in the linear regime.

Besides the linear instability in driven dissipative systems, much attention were also paid to the research of modulational instability (MI) in nonlinear systems~\cite{Segur2007,Zakharov2009,Zakharov2013,Kibler2015,Biondini2016,
Conforti2016,Mosca2018,Kraych20191}. MI is a typical nonlinear instability discovered firstly in the study of water waves~\cite{Benjamin1967},
and may apply to non-dissipative and non-driven systems, where a plane wave of finite amplitude may undergo an instability and lose its energy to  sideband components, resulting in a nonlinear modulation of the plane wave. Usually, MI is considered as wave dynamics problems for  conservative nonlinear systems, where wave envelopes are controlled typically by cubic nonlinear Schr\"{o}dinger~(NLS) equation, with local and attractive Kerr nonlinearities. For such systems, the existence of MI is thought to be the major reason for the formation of solitons
(see Refs.~\cite{Segur2007,Zakharov2009,Zakharov2013,Kibler2015,Biondini2016,Conforti2016,Mosca2018,Kraych20191,Reece2007,Solli,
Nguyen2017,Kraych20192,Benjamin1967,Kamchatnov1997,Kartashov2011,Kivshar2006,Agrawal2007,Pethick2008} and references cited therein).

In recent years, considerable efforts were made on the study of
MI in conservative nonlinear systems described by cubic NLS equations with nonlocal Kerr nonlinearity. It was found that MI may occur in such systems even the Kerr nonlinearity is repulsive, or has both repulsive and attractive parts~\cite{Krolikowski2001,Wyller2002,Krolikowski2004,
Doktorv2007,Henkel2010,Esbensen2011,Tiofack2015,Maucher2016,
Maucher2017}, which provides the possibility to realize spontaneous symmetry breaking~\cite{Malomed2013} and generate spatially extended, ordered structures in nonlocal nonlinear media. Various self-organized patterns were found~\cite{Tiofack2015,Maucher2016,Maucher2017,Mottl2012,Labeyrie2014,
ZhangYC2018}, especially the ones of atomic density formed in Rydberg-dressed~\cite{Henkel2010,Cinti2010,Cinti2014,Henkel2012,Hsuch2012,
Hsuch2017,Li2018} and dipolar~\cite{Saito2009,Lu2015,Kadau2016,Wachtler2016,Xi2018,Zhang2019} Bose-Einstein condensates.

On the other hand, in the past two decades a large amount research works were focused on the investigation of cold Rydberg atomic gases~\cite{Gallagher2006,Saffman2010} working under condition of electromagnetically induced transparency (EIT). EIT is an important quantum destruction interference effect occurring typically in resonant three-level atomic systems, by which the absorption of a probe laser field can be largely eliminated by a control laser field~\cite{Fleischhauer2005}. Due to strong, long-range atom-atom interaction~(also called Rydberg-Rydberg interaction),
such systems are desirable nonlinear optical media with strong, nonlocal Kerr nonlinearity if the Rydberg-Rydberg interaction is suitably mapped to photon-photon interaction via EIT~\cite{Fir2016,Mur2016}. In an interesting work, Sevinli {\it et al.}~\cite{Sevincli2011} reported  a self-organized hexagonal optical pattern via a MI of plane-wave probe beam in a cold Rydberg atomic gas with a repulsive Rydberg-Rydberg interaction.

In this work, we propose and analyze a scheme for realizing various self-organized optical structures and their structural phase transition in a cold Rydberg atomic gas via a Rydberg-EIT~\cite{Mohapatra2007,Pritchard2010}. By exploiting a microwave dressing (i.e., a microwave field couples two electrically excited Rydberg states)~\cite{Tana2011,Sedlacek2012,Yu2013,Maxwell2013,Petrosyan2014,Pohl2014,Li2014,
Li2015,Rao2014,Adams2014,Liu2015,Thompson2017,Votg2018,Vogt2019,Jing2020}, we show that the nonlocal Kerr nonlinearity of the Rydberg gas (which has only a repulsive Rydberg-Rydberg interaction in the absence of the microwave field) is significantly modified, and its strength and sign can be tuned actively. Based on such nonlocal Kerr nonlinearity, we demonstrate that a homogeneous (plane wave) state of probe laser field can undergo MI and hence spontaneous symmetry breaking, which may result in the formation of various ordered optical patterns.

Through detailed analytical and numerical analysis, we find that a homogeneous state of the probe field is firstly transited into a hexagonal lattice pattern (which is the only lattice pattern when the microwave dressing is absent). Interestingly, this hexagonal lattice pattern may undergo a structural phase transition and develop into several types of square lattice patterns when the microwave field is applied and its strength is increased. Moreover, as a outcome of the MI the formation of nonlocal spatial optical solitons is also possible by a suitable choice of system parameters. Different from the results reported before, the optical patterns and nonlocal optical solitons found here can be flexibly manipulated via the adjustment of the effective probe-field intensity, nonlocality degree of the Kerr nonlinearity, and the strength of the microwave field. Our study opens a way for actively controlling the self-organization and structural phase transition of optical patterns through microwave-dressing on Rydberg gases, which are not only of fundamental interest, but also useful for potential applications in optical information processing and transmission.

The remainder of the article is arranged as follows. In Sec.~\ref{sec2}, we describe the physical model, discuss the modification and enhancement of the Kerr nonlinearity contributed by the microwave field, and derive the nonlinear envelope equation of the probe field. In Sec.~\ref{sec3}, we consider the MI of a plane-wave state, investigate the formation and structural phase transitions of optical patterns controlled by the microwave field, effective probe-field intensity, the nonlocal Kerr nonlinearity and its nonlocality degree. The result on the formation of nonlocal spatial optical solitons is also presented. The last section (Sec.~\ref{sec4}) gives a summary of our main research results.

\section{Physical model, nonlinear envelope equation, and enhanced Kerr nonlinearity}\label{sec2}

\subsection{Physical model}\label{sec21}

\begin{figure}[t]
\centering
\includegraphics[width=0.5\textwidth]{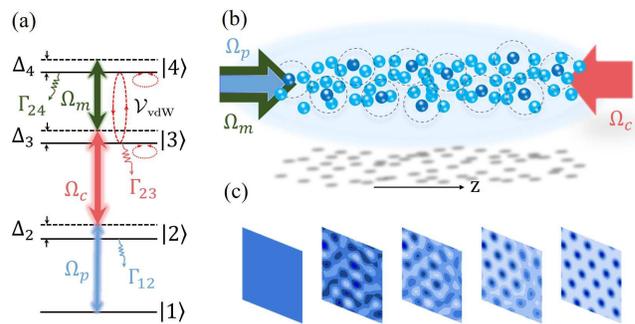}
\caption{\footnotesize Schematics of the model.
(a)~Ladder-type four-level atomic configuration for realizing the microwave-dressed Rydberg-EIT. Here, the weak probe laser field~(blue), strong control laser field~(red), and strong microwave field~(green) with
half Rabi frequencies ${\Omega}_{p}$, $\Omega_c$, and $\Omega_m$ drive the transitions $|1\rangle\leftrightarrow|2\rangle$, $|2\rangle\leftrightarrow|3\rangle$, and $|3\rangle\leftrightarrow|4\rangle$, respectively;
States $|1\rangle$ and $|2\rangle$ are respectively ground and excited states, both  $|3\rangle$ and $|4\rangle$ are highly excited Rydberg states; $\Delta_2$, $\Delta_3$, and $\Delta_4$ are respectively the one-, two-, and three-photon detunings; $\Gamma_{12}$, $\Gamma_{23}$, and $\Gamma_{24}$ are the spontaneous emission decay rates from $|2\rangle$ to $|1\rangle$,  $|3\rangle$ to $|2\rangle$ and
$|4\rangle$ to $|2\rangle$, respectively.
Two Rydberg atoms locating respectively at position ${\bf r}$ and ${\bf r}'$ interact through van der Waals potential $\hbar {\mathcal V}_{\rm vdw}^{l}({\bf r}^{\prime}-{\bf r})$ ($l=s,d,e$; see text).
(b)~Possible experimental geometry, where small solid circles denote atoms and large dashed circles denote Rydberg blockade spheres.
(c)~Emergence of an optical pattern via modulation instability.
}\label{Fig1}
\end{figure}

We consider an ensemble of lifetime-broadened four-level atomic gas with an ladder-type level configuration, shown schematically in Fig.~{\ref{Fig1}}{\color{blue}(a)}.
Here, the weak probe laser field with central angular frequency $\omega_{p}$, wavevector ${\bf k}_{p}$, and half Rabi frequency
$\Omega_p$ drives the transition from atomic ground state $|1\rangle$ to  intermediate state $|2\rangle$, and the strong control laser field with
central angular frequency $\omega_c$, wavevector ${\bf k}_c$, and
half Rabi frequency $\Omega_c$ drives the transition $|2\rangle$ to the first highly excited Rydberg state $|3\rangle$. This ladder-type three level EIT is dressed by a microwave field with central angular frequency
$\omega_m$, wavevector ${\bf k}_m$, and half Rabi frequency
$\Omega_m$, which couples the transition between the Rydberg state $|3\rangle$ and another Rydberg state $|4\rangle$.
The total electric fields acting in the atomic system can be written as ${\bf E(r},t)=\sum_{j}{\bf e}_{j}{\cal E}_{j}e^{i({\bf k}_{j}\cdot{\bf r}-\omega_{j}t)} +{\rm  H.c.}$, with ${\bf e}_{j}$ and ${\cal E}_{j}$ respectively the polarization unit vector and the envelope of $j$-th laser field ($j=p,c,m$).  $\Delta_2$, $\Delta_3$, and $\Delta_4$ are respectively the one-, two-, and three-photon detunings; $\Gamma_{12}$, $\Gamma_{23}$, and $\Gamma_{24}$  are the spontaneous emission decay rates from $|2\rangle$ to $|1\rangle$,  $|3\rangle$ to $|2\rangle$, and $|4\rangle$ to $|2\rangle$, respectively.
The microwave field employed here is to realize a microwave-dressed Rydberg-EIT~\cite{Tana2011,Sedlacek2012,Yu2013,Maxwell2013,Petrosyan2014,Pohl2014,Li2014,Li2015,
Rao2014,Adams2014,Liu2015,Thompson2017,Votg2018,Vogt2019,Jing2020} and thus to modify the Rydberg-Rydberg interaction,
which, in turn, can manipulate the interaction strength and sign for the photons in the probe field and hence realize self-organized optical structures not discovered before.

The dynamics of the system is controlled by the Hamiltonian $\hat{H}=\mathcal{N}_a\int d^3{\bf r}\hat{\mathcal{H}}({\bf r},t)$, where $\hat{\mathcal{H}}({\bf r},t)$ is Hamiltonian density and $\mathcal{N}_a$ is atomic density. Under the electric-dipole and rotating-wave approximations, the Hamiltonian density in interaction picture reads
\begin{align}\label{Hamiltonian}
\hat{\mathcal{H}}\equiv\hat{\mathcal{H}}_1+\hat{\mathcal{H}}_{{\rm vdW}},
\end{align} \noindent
where Hamiltonian $\hat{\mathcal{H}}_1$ describes unperturbed atoms as well as the interaction between the atoms and the laser fields,  $\hat{\mathcal{H}}_{{\rm vdW}}$ describes the Rydberg-Rydberg interaction, respectively given by
{\small
\begin{subequations}
\begin{align}
&\hat{\mathcal{H}}_1=-\hbar\sum_{\alpha=2}^4\Delta_{\alpha}\hat{S}_{\alpha\alpha} -\hbar(\Omega_p \hat{S}_{12}+\Omega_c
\hat{S}_{23}+\Omega_m \hat{S}_{34}+{\rm H.c.}),\\
&\hat{\mathcal{H}}_{{\rm vdW}}=\hbar\mathcal{N}_a\int d^3{ r}^\prime\bigg\{\sum_{\alpha=3,4}\hat{S}_{\alpha\alpha}({\bf
r}^{\prime},t)\mathcal{V}_{\alpha\alpha}^s({\bf r}^\prime-{\bf r})\hat{S}_{\alpha\alpha}({\bf r},t)\notag\\
&\qquad\,+\mathcal{V}_{34}^d({\bf r}^\prime-{\bf r})\left[\hat{S}_{33}({\bf
r^\prime},t)\hat{S}_{44}({\bf r},t)+\hat{S}_{44}({\bf r^\prime},t)\hat{S}_{33}({\bf r},t)\right]\notag\\
&\qquad\,+\mathcal{V}_{34}^e
({\bf r}^\prime-{\bf r})\left[\hat{S}_{43}({\bf
r^\prime},t)\hat{S}_{34}({\bf r},t)+\hat{S}_{34}({\bf r^\prime},t)\hat{S}_{43}({\bf r},t)\right]\bigg\}.
\end{align}
\end{subequations}}\noindent
Here $d^3 r'=dx' dy' dz'$;
$\hat{S}_{\alpha\beta}=|\beta\rangle \langle \alpha|\exp\{i[({\bf k}_{\beta}-{\bf k}_{\alpha})\cdot{\bf r}-(\omega_{\beta}-\omega_{\alpha}+\Delta_{\beta}-\Delta_{\alpha})t]\}$
is the transition operator satisfying the commutation relation
$[\hat{S}_{\alpha\beta}({\bf
r},t),\hat{S}_{\alpha^\prime\beta^\prime}({\bf r}^\prime,t)]=\mathcal{N}_a^{-1}
\delta ({\bf r}-{\bf r}{^\prime})(\delta_{\alpha\beta'}\hat{S}_{\alpha'\beta}({\bf r},t)-\delta_{\alpha'\beta}\hat{S}_{\alpha\beta'}({\bf r'},t))$;
the one-, two-, and three-photon detuings are respectively given by
$\Delta_2=\omega_p-(\omega_2-\omega_1)$, $\Delta_3= \omega_c+\omega_p-(\omega_3-\omega_1)$, and
$\Delta_4=\omega_c+\omega_p+\omega_m-(\omega_4-\omega_1)$, with
$E_{\alpha}=\hbar\omega_{\alpha}$ the eigenenergy of the atomic state $|\alpha\rangle$.
The half Rabi frequencies of the probe, control, and microwave fields are, respectively, $\Omega_p=({\bf e}_{p}\cdot{\bf p}_{21}){\cal
E}_{p}/\hbar$, $\Omega_c=({\bf e}_{c}\cdot{\bf p}_{32}){\cal E}_{c}/\hbar$, and $\Omega_m=({\bf e}_{m}\cdot{\bf p}_{43}){\cal
E}_{m}/\hbar$, with ${\bf p}_{\alpha\beta}$  the electric dipole matrix element associated with the transition between the states $|\alpha \rangle$ and $|\beta\rangle$.

The Hamiltonian density $\hat{\mathcal{H}}_{{\rm vdW}}$ is the contribution by the Rydberg-Rydberg interaction, which contains
four parts, represented by $\mathcal{V}_{33}^s$, $\mathcal{V}_{44}^s$, $\mathcal{V}_{\rm 34}^d$, and $\mathcal{V}_{\rm 34}^e$,
respectively; the term $\mathcal{V}_{\rm 33}^s=-C_{33}^{s}/|{\bf r}'-{\bf r}|^6$\,\,  ($\mathcal{V}_{44}^s=-C_{44}^{s}/|{\bf r}'-{\bf r}|^6$)
describes the van der Waals interaction between the two
atoms located respectively at positions ${\bf r}'$ and ${\bf r}$
and excited to the same Rydberg state $|3\rangle$\, ($|4\rangle$); the term $\mathcal{V}_{\rm 34}^{d}=-C_{34}^{d}/|{\bf r}'-{\bf r}|^6$\,\,
($\mathcal{V}_{\rm 34}^e=-C_{34}^{e}/|{\bf r}'-{\bf r}|^3$)
describes the direct non-resonant van der Waals interaction
(resonant exchange dipole-dipole interaction)
between the two atoms
excited to different Rydberg states (i.e. $|3\rangle$ and $|4\rangle$). Here $C_{\alpha\beta}^{l}$   ($\{\alpha\beta\}=\{33,44,34\}$; $l=s,d,e$) are dispersion parameters~\cite{Petrosyan2014,Li2014,Li2015}.

The time evolution of the atoms in the system is governed by the optical
Bloch equation
\begin{equation}\label{Bloch0}
\frac{\partial\rho}{\partial t}=-\frac{i}{\hbar} \left[{\hat{ H}},\rho\right]-\Gamma\left[\rho\right],
\end{equation}
where $\rho ({\bf r},t)=\langle \hat{S}({\bf r},t)\rangle$~\cite{note0} is a $4\times 4$ density matrix (with density matrix elements $\rho_{\alpha\beta}({\bf r},t)=\langle \hat{S}_{\alpha\beta} ({\bf r},t)\rangle$; $\alpha,\beta=1,2,3,4$)
describing the atomic population and coherence, $\Gamma$ is a $4\times 4$ relaxation matrix describing the spontaneous emission
and dephasing. Explicit expressions of $\rho_{\alpha\beta}({\bf r},t)$ are  presented in Appendix~\ref{app1}.

The propagation of the probe field is controlled by Maxwell equation, which under paraxial and slowly-varying envelope approximations is reduced into~\cite{Mur2016}
\begin{align}\label{Maxwell}
  i\left(\frac{\partial}{\partial z}+\frac{1}{c}\frac{\partial}{\partial
  t}\right)\Omega_p+\frac{c}{2\omega_p}\nabla_{\perp}^2\Omega_p
  +\kappa_{12}\rho_{21}=0,
\end{align}\noindent
where $\nabla_{\perp}^2=\partial^2/\partial x^2+\partial^2/\partial y^2$ describes diffraction, $\kappa_{12}=\N_a\omega_p|({\bf e}_p\cdot{\bf p}_{12})|^2/(2\epsilon_0c\hbar)$ is a parameter describing
the coupling between the atoms and the probe field,
and $c$ is the light speed in vacuum.
Without loss of generality, we assume the probe field propagates along $z$ direction, i.e., ${\bf k}_p=(0,0,\omega_p/c)$; to suppress Doppler effect, the microwave field is along the $z$ direction but the control field is along the negative $z$ direction [i.e., ${\bf k}_m=(0,0,\omega_m/c)$ and ${\bf k}_c=(0,0,-\omega_c/c)$]. A possible experimental arrangement is given in Fig.~\ref{Fig1}(b).

Note that the physical model described above is valid for any microwave-dressed Rydberg atomic gas. But for latter calculations where numerical values of the system are needed, we take cold $^{87}$Rb atomic gas~\cite{Petrosyan2014} (which has density-density interaction in the absence of the microwave field) as a realistic example. The assigned atomic levels in Fig.~\ref{Fig1}(a) are $|1\rangle=|5 S_{1/2}\rangle$, $|2\rangle=|5P_{3/2}\rangle$,  $|3\rangle=|nS_{1 / 2}\rangle$, and $|4\rangle=|n P_{3 / 2}\rangle$. For example, for principal quantum number $n=60$, the dispersion parameters are
$C_{33}^{s}=- 2\pi \times 140\,{\rm GHz}~\mu {\rm {m^6}}$~(repulsive  interaction), $C_{44}^{s}=2\pi \times 295  \,{\rm GHz}~\mu {\rm
{m^6}}$~(attractive interaction), $C_{34}^{d}=-2\pi \times 3 \,{\rm GHz}~\mu {\rm {m^6}}$ (repulsive interaction), and
$C_{34}^{e}=-2\pi \times  3.8 \,{\rm GHz}~\mu {\rm {m^3}}$~(repulsive  interactions)~\cite{Petrosyan2014,Rao2014,note1}, respectively.
Typical system parameters are chosen as follows:
$\Delta_2=3.17\times 10^2$\,MHz,
$\Delta_3=15.3$\,MHz,
$\Delta_4=1.32$\,MHz;
$\Gamma_{12}=2\pi \times 6.1$\,MHz,
$\Gamma_3=\Gamma_{4}=2\pi \times 1.67\times 10^{-2}$\,MHz;
$\Omega_{c}=20\,{\rm MHz}$;
$\N_a=1.0\times 10^{11}\,{\rm cm^{-3}}$.

We stress that although the Bloch Eq.~(\ref{Bloch0}) is for the evolution of one-body density-matrix elements $\rho_{\alpha\beta}({\bf r},t)$, it involves two-body density-matrix elements $\rho_{\alpha\beta,\mu\nu}({\bf r^{\prime},r},t)= \langle
\hat{S}_{\alpha\beta}({\bf r^{\prime}},t) \hat{S}_{\mu\nu}({\bf r},t)\rangle$ due to the Rydberg-Rydberg interaction; furthermore, the equation of motion for $\rho_{\alpha\beta,\mu\nu}({\bf r^{\prime},r},t)$ involves three-body density-matrix elements $\rho_{\alpha\beta,\mu\nu,\gamma\delta}({\bf r}^{\prime\prime},{\bf r}^{\prime}, {\bf r},t)=\langle
\hat{S}_{\alpha\beta}({\bf r}^{\prime\prime},t) \hat{S}_{\mu\nu}({\bf r}',t)\rangle \hat{S}_{\gamma\delta}({\bf r},t)\rangle$, and so on. Thus an effective approach for solving such a hierarchy of infinite equations involving many-atom correlations is needed.

\subsection{Enhanced Kerr nonlinearity by the microwave dressing}\label{sec22}

We first consider the modification of Kerr nonlinearity of the system induced by the microwave field based on the physical model described above. For simplicity, we assume that the control and microwave fields are strong enough, so that they are not depleted during the propagation of the probe field. Since the probe field is weak, a perturbation expansion can be applied to solve the Maxwell-Bloch (MB) equations (\ref{Bloch0}) and (\ref{Maxwell}) by taking $\Omega_p$ as a small expansion parameter. Generalizing the approach developed in Refs.~\cite{Bai2016,Zhang2018,Bai2019,Bai2020}, where MB equations for Rydberg atomic gases without microwave dressing are solved beyond mean-field approximation in a self-consistent and effective way, we can obtain the solutions of the Bloch Eq.~(\ref{Bloch0}) using the perturbation expansion up to third-order approximation. In particular, the result of the one-body density matrix element $\rho_{21}$ can be obtained analytically (see the Appendix~\ref{app1} for detail).

With the expression of $\rho_{21}$ and the definition of probe-field susceptibility, i.e. $\chi= \mathcal{N}_a({\bf e\cdot p}_{12})\rho_{21}/(\epsilon_0\mathcal{E}_p)$, it is easy to obtain the optical susceptibility of the probe field, which reads
\begin{equation}
\chi=\chi^{(1)}+\left(\chi_{\rm loc}^{(3)}+\chi_{\rm nloc}^{(3)}\right)|\mathcal{E}_p|^2,
\end{equation}
where $\chi^{(1)}$ is the linear susceptibility;
$\chi_{\rm loc}^{(3)}$ and $\chi_{\rm nloc}^{(3)}$ are local and nonlocal third-order nonlinear (Kerr) susceptibilities, originated respectively from non-zero two-photon detuning (i.e. $\Delta_3\neq 0$)~\cite{Bai2016,Zhang2018,Bai2019,Bai2020,Wang2001,Chen2014}) and from the Rydberg-Rydberg interaction in the system.
Expressions of $\chi^{(1)}$, $\chi_{\rm loc}^{(3)}$, and $\chi_{\rm nloc}^{(3)}$ are given in Appendix~\ref{app3}; see
Eqs.~(\ref{chi3s1}), (\ref{chi3s2}), and (\ref{chi3n}), respectively.
Using the system's parameters given at the final part of the last subsection and taking
$\Omega_m=18$\,MHz, we obtain
$\chi^{(3)}_{\rm loc} \approx (5.08+0.012i)\times 10^{-11}\, {\rm m^{2}/V^2}$,
$\chi^{(3)}_{\rm nloc}\approx (3.05+0.022i)\times 10^{-8}\,
{\rm m^{2}/V^2}$.
We see the imaginary parts of $\chi^{(3)}_{\rm loc}$ and $\chi^{(3)}_{\rm nloc}$ are much smaller than their real parts,
which is due to the EIT effect contributed by the control field; moreover, the nonlocal Kerr nonlinearity is three orders of magnitude larger than the local one, which is due to the strong Rydberg-Rydberg interaction together with the microwave dressing.

\begin{figure}
\centering
\includegraphics[width=0.48\textwidth]{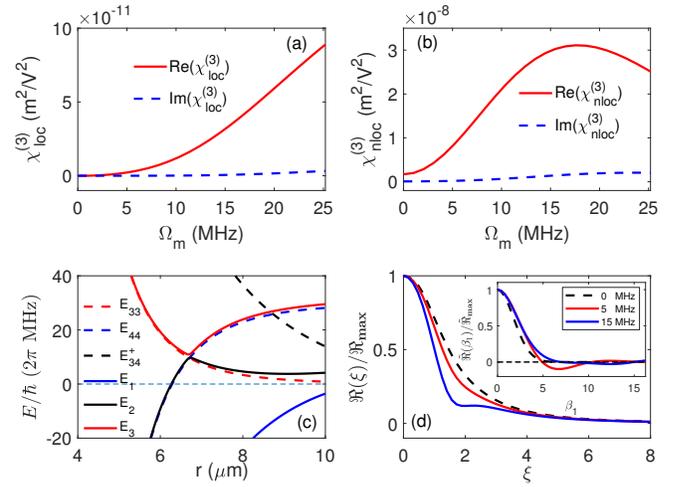}
\caption{\footnotesize
Kerr nonlinearity enhancement, interaction potential energy, and normalized nonlinear response function $\Re$ of the probe field in the presence of the microwave dressing.
(a)~Real part Re($\chi^{(3)}_{\rm loc}$)~(solid red line) and  imaginary part Im($\chi^{(3)}_{\rm loc})$~(dotted blue line) of the local nonlinear susceptibility $\chi_{\rm
loc}^{(3)}$ as a function of the half Rabi frequency $\Omega_m$ of the microwave field.
(b)~The same as (a) but for the nonlocal nonlinear susceptibility $\chi_{\rm nloc}^{(3)}$.
(c)~Potential-energy curves $E_{1}$ (solid blue line), $E_2$ (solid black line), and $E_3$ (solid red line) of two Rydberg atoms
as functions of the interatomic distance $r= |{\bf r'-r}|$ for $\Omega_m=10~$MHz and $\Delta=13.98$ MHz; $E_{33}$ (dashed blue line), $E_{44}$ (dashed black line), and $E_{34}^{+}$ (dashed red line) are for the case without the microwave field.
(d)~Normalized response function $\Re/\Re_{\rm max}$ as a function of the dimensionless coordinate $\xi= x/R_0$ ($R_0$ is the typical transverse beam radius of the probe field) with $\Omega_m=0$~(dotted black line), 5~MHz~(solid red line), and 15~MHz~(solid blue line), respectively.
Inset: the normalized response function $\tilde{\Re}/\tilde{\Re}_{\rm max}$ in momentum space (i.e. the Fourier transformation of
$\Re/\Re_{\rm max}$)
as a function of the dimensionless wavenumber
$\beta_1=R_0 k_x$ ~($k_x$ is the dimensional wavenumber  in $x$ direction) with $\Omega_m=0$, 5, and 15\,MHz, respectively.
}
\label{Fig2}
\end{figure}

It is helpful to reveal how the microwave-dressing modify the Kerr effect
of the system. Fig.~\ref{Fig2}(a) shows
the real part Re($\chi^{(3)}_{\rm loc}$)~(solid red line) and  imaginary part Im($\chi^{(3)}_{\rm loc})$~(dotted blue line) of the local nonlinear susceptibility $\chi_{\rm loc}^{(3)}$ as a function of the half Rabi frequency $\Omega_m$ of the microwave field. Shown in Fig.~\ref{Fig2}(b)
is same as that in Fig.~\ref{Fig2}(a) but for the nonlocal nonlinear susceptibility $\chi_{\rm nloc}^{(3)}$.
From the figure we see that the nonlinear optical susceptibilities have two evident features:
(i)~Both the real parts of $\chi_{\rm
loc}^{(3)}$ and $\chi_{\rm nloc}^{(3)}$ are much larger than the corresponding imaginary parts, contributed by the EIT effect;
(ii)~In the value range of $\Omega_m$ taken here,
${\rm Re(\chi_{loc}^{(3)})}$ is an increasing function; however,
${\rm Re(\chi_{\rm nloc}^{(3)}}$) increases firstly, then arrives a maximum at some value of $\Omega_m$, and decreases when $\Omega_m$ is increased further. At the point of the maximum, where
$\Omega_m\approx 18$ MHz, ${\rm Re}(\chi_{\rm nloc}^{(3)})
\approx 3.05\times 10^{-8}\,{\rm m}^2/{\rm V}^2$, which is 15 times larger than the case without the microwave field
[${\rm Re}(\chi_{\rm nloc}^{(3)})\approx 0.2\times 10^{-8}~{\rm m^2/V^2}$ for $\Omega_m=0$]. Thus, microwave-dressing can be used to modify the Kerr effect of the system greatly.

To support the above conclusion, a calculation is carried out for the interaction potential  of two Rydberg atoms
located respectively at positions ${\bf r}$ and
${\bf r}'$, which may occupy in the Rydberg states $|3\rangle$ and $|4\rangle$. In the absence of the microwave field (i.e., $\Omega_m=0$), the basis set of the such two-atom system consists of states
$|33\rangle= |3_13_2\rangle$,
$|44\rangle= |4_14_2\rangle$, and
$|34_\pm \rangle= 1/\sqrt{2}(|3_14_2\rangle\pm|3_24_1\rangle )$,
with the subscript 1 and 2 representing atom 1 and 2, respectively.
The (bare state) eigen energies of the system are $E_{33}=-\hbar C_{33}^s/{\bf |r'-r|}^6$, $E_{44}=2\hbar\Delta -\hbar C_{44}^s/{\bf |r'-r|}^6$, $E_{34}^{+}=\hbar \Delta+ \hbar C_{34}^e/|{\bf r'-r}|^3$, and $E_{34}^{-}=\hbar \Delta- \hbar C_{34}^d/|{\bf r'-r}|^6$, with $\Delta=\Delta_3-\Delta_4=13.98$ MHz.
Since antisymmetric state $|34_{-}\rangle$ is nearly not coupled to  laser field, one can disregard it if the microwave field is present (i.e., $\Omega_m\neq 0$). Then, the Hamiltonian in the two-atom basis set $\{|33\rangle$, $|34_{+}\rangle$, $|44\rangle\}$ takes the form
\begin{align}\label{HTA}
  \mathcal{H}=\hbar\left(
  \begin{array}{ccc}
    -\frac{C_{33}^s}{|{\bf r'-r}|^6}&\sqrt{2}\Omega_m&0\\
    \sqrt{2}\Omega_m&\Delta+\frac{C_{34}^e}{|{\bf r'-r}|^3}&\sqrt{2}\Omega_m\\
    0&\sqrt{2}\Omega_m&2\Delta-\frac{C_{44}^s}{|{\bf r'-r}|^6}\\
  \end{array}
  \right).
\end{align}
After diagonalization, we can obtain the energies $E_{1}$, $E_2$, and $E_3$ of the Hamiltonian (\ref{HTA}).
Potential-energy curves of $E_{1}$, $E_2$, and $E_3$ as functions of the interatomic separation $r= |{\bf r'-r}|$ for $\Omega_m=10~$MHz are shown in Fig.~\ref{Fig2}(c). For comparison,
the bare potential-energy curves $E_{33},\, E_{44}$, and $E_{34}^+$ (for $\Omega_m=0$) are also shown. We see that, compared with the case without the microwave field, the potential-energy curves are modified largely by the introduction of the microwave field, especially for small interatomic separation $r$. The reason is that the microwave dressing brings a coupling between the Rydberg states $|3\rangle$ and $|4\rangle$, and thereby a modification of the Rydberg-Rydberg interaction. It is the use of the microwave dressing that brings the significant change and enhancement of the nonlocal Kerr nonlinearity.

\subsection{Nonlinear envelope equation and the property of nonlinear response function}\label{sec23}

We now derive the envelope equation which controls the dynamics of the probe field. By substituting the solution of $\rho_{21}$ into
the Maxwell Eq.~(\ref{Maxwell}) and making a local approximation along the $z$ direction on the nonlocal nonlinear response function~(see the Appendix~\ref{app3}),
we obtain the following three-dimensional (3D) nonlocal nonlinear Schr\"odinger (NNLS) equation
{\small
\begin{align}
& i\frac{\partial \Omega_p }{\partial z}+\frac{c}{2\omega_p}\nabla_{\perp}^{2}\Omega_p + {W_1}|\Omega_p{|^2}\Omega_p\notag\\
& \qquad+\int d^2 {r'}
G({\bf r_{\perp}'-r_{\perp}})|\Omega_p({\bf r_{\perp}'},z)|^2\Omega_p({\bf r_{\perp}},z)=0,\label{NNLS1}
\end{align}}\noindent
with ${\bf r}_{\perp}=(x,y)$,  $d^2{r'}=dx' dy'$.
The third and forth terms on the left hand side of this equation describe two types of self-phase modulations of the probe field, contributed respectively by the local Kerr nonlinearity (originated from non-zero two-photon detuning (i.e., $\Delta_3\neq 0$)~\cite{Bai2016,Zhang2018,Bai2019,Bai2020,Wang2001,Chen2014}) and the nonlocal Kerr nonlinearity (originated from the Rydberg-Rydberg interaction). In the integral of the forth term of the NNLS equation, $G$ is a reduced nonlinear response function, taking the form
$G({\bf r_{\perp}'-r_{\perp}})=\sum_{\alpha=3,4}G_{\alpha\alpha}^s({\bf r_{\perp}'-r_{\perp}})+\sum_{l={d,e}}G_{34}^l({\bf r_{\perp}'-r_{\perp}})$,
with
$G_{\alpha\beta}^l({\bf r_{\perp}'-r_{\perp}})=\int dz'R_{\alpha\beta}^l({\bf r'-r})$
($\{\alpha\beta\}=\{33,44,34\}$; $l=s,d,e$). Explicit expressions
of the matrix elements of nonlinear response function
$R_{\alpha\beta}^{l}({\bf r}^\prime-{\bf r})$ and local nonlinear coefficient $W_1$ are given in the Appendix~\ref{app3}~[see Eq.~(\ref{W1}) and Eqs.~(\ref{eqb5}), respectively]. Due to the microwave dressing,
the nonlocal Kerr nonlinearity consists of four parts;  the first~(second) part $G_{33}^s({\bf r_{\perp}'-r_{\perp}})$
[$G_{44}^s({\bf r_{\perp}'-r_{\perp}})$] is contributed by the interaction of the atoms lying in the same Rydberg state $|3\rangle$~($|4\rangle$);
the third (forth) part $G_{34}^d({\bf r_{\perp}'-r_{\perp}})$~[$G_{34}^e({\bf r_{\perp}'-r_{\perp}})$] is contributed by the interaction of the atoms lying in the different Rydberg states $|3\rangle$ and $|4\rangle$.

For the convenience of later discussions and numerical calculations, we rewrite the 3D NNLS Eq.~(\ref{NNLS1}) into the non-dimensional form
\begin{align}\label{NNLS2}
&i\frac{\partial u}{\partial s} + \tilde{\nabla}_{\perp}^2u+\int d^2\zeta'
\Re(\vec{\zeta}'-\vec{\zeta})|u(\vec{\zeta}',s){|^2}
u(\vec{\zeta},s)=0,
\end{align}
with $s = z/(2{L_{\rm diff}})$, $u=\Omega_p/U_0$,
$\tilde{\nabla}_{\perp}^2=\partial^2/\partial\xi^2
+\partial^2/\partial\eta^2$,
$\vec{\zeta}= (\xi,\eta)=(x,y)/{R_0}$, and
$d^2 {\zeta}^{\prime}= d\xi^{\prime}d \eta^{\prime}$.
Here ${L_{\rm diff}} = {\omega _p}R_0^2/c$ is the typical diffraction length, which is 1.61 mm in our system;
$U_0$ is the typical Rabi frequency of the probe field;
$R_0$ is the typical beam radius of the probe field;
the non-dimensional nonlinear response function is defined by
$\Re(\vec{\zeta}'-\vec{\zeta})=2L_{\rm diff}U_0^2R_0^2 G[{(\vec{\zeta}'-\vec{\zeta})}R_0]$. Note that in writing Eq.~(\ref{NNLS2})
we have neglected the term related to $W_1$ because the
local Kerr nonlinearity is much smaller than the nonlocal one~\cite{Bai2019}.

The property of the nonlocal Kerr nonlinearity of the system is characterized by the nonlinear response function $\Re(\vec{\zeta})$.
Comparing with the case without the microwave field ($\Omega_m=0$), $\Re(\vec{\zeta})$ is largely modified and can be  manipulated by the use of the microwave field ($\Omega_m\neq 0$). To demonstrate this, the normalized response function $\Re/\Re_{\rm max}$ as a function for $\xi= x/R_0$ is shown in Fig.~\ref{Fig2}(d), where the dotted black line, solid red line, and solid blue line are for $\Omega_m=0$, $\Omega_m=5$~MHz, and $\Omega_m=$15~MHz, respectively.
We see that, due to the role played by the microwave field, the shape of
$\Re/\Re_{\rm max}$ is changed significantly. Especially, for a larger  microwave field, the $\Re/\Re_{\rm max}$ curve
becomes negative for the small value of $\xi$ (not shown here), consistent with the result  obtained in Ref.~\cite{Petrosyan2014}.
Plotted in the inset of the figure is the normalized response function in momentum space $\tilde{\Re}/\tilde{\Re}_{\rm max}$
(i.e., the Fourier transformation of $\Re/\Re_{\rm max}$)
as a function of the non-dimensional wavenumber
$\beta_1=R_0 k_x$ ~($k_x$ is the wavenumber in $x$ direction) with $\Omega_m=0$, 5, and 15\,MHz, respectively.
One sees that $\tilde{\Re}/\tilde{\Re}_{\rm max}$ has only one change in sign for $\Omega_m=0$; however, more changes in sign arise when $\Omega_m$ takes nonzero values. Such behavior of $\tilde{\Re}/\tilde{\Re}_{\rm max}$ is due to the joint action by the nonlocal Kerr nonlinearity and the microwave field, through which the MI of the plane-wave probe field may occur~(see the next section).

Except for the significant dependence on the microwave field $\Omega_m$,
the property of the response function depends also on another parameter, i.e., the nonlocality degree of the Kerr nonlinearity, defined by
\begin{equation}\label{NLD}
\sigma\equiv {R_b}/{R_0},
\end{equation}
where $R_b$ is the radius of Rydberg blockade sphere, given by
$R_b=|C_{33}^{s}/\delta_{\rm EIT}|^{1/6}$~\cite{Saffman2010,Fir2016,Mur2016}, with
$\delta_{\rm EIT}$ the width of EIT transparency window. One has
$\delta_{\rm EIT}=|\Omega_c|^2/\gamma_{21}$ for
$\Delta_2=0$, and $\delta_{\rm EIT}=|\Omega_c|^2/\Delta_2$
for $\Delta_2\gg\gamma_{21}$. With the system parameters used here,
we have $R_b\approx 8.34 ~\mu{\rm m}$.
In the next section, we shall show that the structural phase transitions of the optical patterns of the system depend strongly not only on the microwave field $\Omega_m$ but also on the nonlocality degree $\sigma$ of the Kerr nonlinearity.

\section{Modulational instability, emergence of optical patterns and solitons}\label{sec3}

\subsection{Modulation instability}\label{sec31}
MI is a nonlinear instability of constant-amplitude continuous waves  under long-wavelength perturbations, occurring in a variety of contexts where Kerr nonlinearity is attractive and local~\cite{Zakharov2009,Biondini2016}; it can also arise in systems with repulsive but nonlocal Kerr nonlinearity when the perturbations have both long~\cite{Krolikowski2001,Krolikowski2004} and short~\cite{Maucher2016,Maucher2017,Sevincli2011} wavelengths. To explore the MI in the our system, we consider the MI of the plane-wave solution of the NNLS Eq.~(\ref{NNLS2}), i.e.,
\begin{equation}\label{PW}
u_{\rm pw}(\vec{\zeta},s)=A_0\exp\left[-is A_0^2\int \Re(\vec{\zeta})d^2 \zeta\right],
\end{equation}
where $A_0$ is a real number.
Since any perturbation can be expanded as a superposition of many Fourier modes, we make the MI analysis of the plane wave by taking only a periodic mode as the perturbation, i.e.,
\begin{eqnarray}\label{PTS}
\tilde{u}(\vec{\zeta},s)=
&&\left[A_0+a_1e^{i\vec{\beta}\cdot \vec{\zeta}+\lambda
s}+a_2^*e^{-i\vec{\beta}\cdot \vec{\zeta}+\lambda^* s}\right]\nonumber\\
&& \times
\exp\left[-is A_0^2\int \Re(\vec{\zeta})d^2\zeta\right],
\end{eqnarray}
where $a_1$ and $a_2$ are small complex amplitudes of the perturbation and $\vec{\beta}= (\beta_1,\beta_2)$\, ($\beta_1\equiv R_0 k_x$, $\beta_2\equiv R_0 k_y$; $k_x$ $k_y$ are wavenumbers in $x$ and $y$ directions, respectively) is non-dimensional 2D wavevector and $\lambda$ is the growth rate of the perturbation, to be determined yet.

Substituting the perturbation solution (\ref{PTS}) into Eq.~(\ref{NNLS2}) and keeping only linear terms of $a_1$ and $a_2$, it is easy to obtain the expression of the growth rate
\begin{align}
\lambda^2=-{\beta}^2\left[\beta^2-2A_0^2\,\tilde{\Re}(\vec{\beta})\right],
\end{align}
where ${\beta}=\sqrt{\beta_1^2+\beta_2^2}$ and $\tilde{\Re}(\vec{\beta})$  is the response function in momentum space [i.e., the Fourier transformation of $\Re(\vec{\zeta})$].

\begin{figure}
\centering
\includegraphics[width=0.49\textwidth]{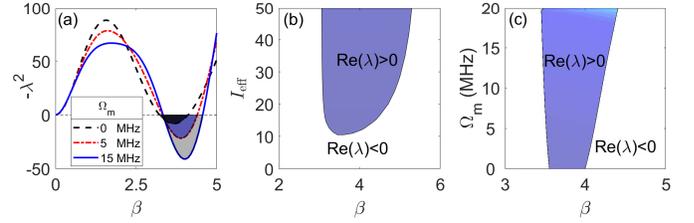}
\caption{\footnotesize Modulation instability and its manipulation for the plane-wave probe field in the Rydberg atomic gas with the microwave dressing.
(a)~$-\lambda^2$ ($\lambda$ is the growth rate) as a function of $\beta=\sqrt{\beta_1^2+\beta_2^2}$  [$\beta_{j}\equiv R_0k_j$; $k_j$ is the wavenumber along $j$th direction ($j=x,y$); $R_0$ is the typical transverse beam radius of the probe field], for
the microwave field $\Omega_m=0$~(dotted black line), 5~MHz~(dashed red line), and 15~MHz~( solid blue line), respectively; shadow regions are ones where MI occurs.
(b)~Real part of the growth rate ${\rm Re}(\lambda$) as  a function of $\beta$ and the effective probe-field intensity $I_{\rm eff}=\alpha A_0^2$\, [with $A_0$ the amplitude of the plane wave and $\alpha=-\int{\Re}(\vec{\zeta})d^2\zeta$\,], for the microwave field $\Omega_m=10$~MHz. The colorful region is the one for ${\rm Re}(\lambda)>0$, where MI occurs.
(c)~${\rm Re}(\lambda$) as a function of $\beta$ and $\Omega_m$ for $I_{\rm eff}=20$; the colorful region is the one where MI occurs.
}\label{Fig3}
\end{figure}

The property of the growth rate $\lambda$ depends on the plane-wave intensity $A_0^2$, the shape of the response function $\tilde{\Re}$ where the microwave field $\Omega_m$ plays an important role.
Shown in Fig.~\ref{Fig3}(a)
is the curve of $-\lambda^2$ as a function of the non-dimensional wavenumber $\beta$ for the microwave field $\Omega_m=0$~(dotted black line), 5~MHz~(dashed red line), and 15~MHz~( solid blue line), respectively. The shadow regions in the figure are ones for ${\rm Re}(\lambda)>0$. That is to say, MI occurs in these shadow regions and hence the plane-wave state of the probe field is unstable. The MI will lead to a symmetry breaking of the system and hence a phase transition to new states. As a result, new optical self-organized structures (or pattern formation) appear in the system (see next section). We note that, different from the cases reported in Refs.~\cite{Krolikowski2001,Krolikowski2004} but similar to those considered in Refs.~\cite{Maucher2016,Maucher2017,Sevincli2011},
the MI in the present system arises for the perturbation of short wavelengths.

To obtain a further understanding of the MI, Fig.~\ref{Fig3}(b) shows
the real part of the growth rate, ${\rm Re}(\lambda$), as  a function of $\beta$ and the effective probe-field intensity
\begin{equation}\label{I}
I_{\rm eff} =\alpha A_0^2
\end{equation}
for $\Omega_m=10$~MHz, where $\alpha=-\int{\Re}(\vec{\zeta})d^2\zeta$ is a parameter characterizing the role by the nonlocal Kerr nonlinearity.
The colorful region in the figure is the one where ${\rm Re}(\lambda)>0$ and hence MI occurs.  Fig.~\ref{Fig3}(c) shows ${\rm Re}(\lambda$) as a  function of $\beta$ and $\Omega_m$ for $I_{\rm eff}=20$, with the colorful region denoting the one where the MI happens.  From these results we see that the MI depends not only on the effective probe-field intensity
$I_{\rm eff}$ but also on the microwave field $\Omega_m$, which provides
ways to manipulate the MI and thereby the emergence of the optical patterns in the system.

\subsection{Pattern formation controlled by the Kerr nonlinearity and the microwave field}\label{sec32}

We now turn to consider the outcome of the MI in the system. Note that
in the absence of the microwave field the system is reduced to a three-level one (i.e. conventional Rydberg-EIT) and the atom-atom interaction Hamiltonian $\hat{\cal H}_{\rm vdw}$ owns only the term
$\hbar\mathcal{N}_a\int d^3{ r}^\prime \hat{S}_{33}({\bf
r}^{\prime},t)\mathcal{V}_{33}^s({\bf r}^\prime-{\bf r})\hat{S}_{33}({\bf r},t)$; however, in the presence of the microwave field, the state $|4\rangle$ may have a significant population and hence it plays an important role for the dynamics of the probe field. In this case $\hat{\cal H}_{\rm vdw}$ owns four terms,
which may be comparable through the tuning of the system parameters.
As a result, the  nonlinear response function $G$ in the envelope Eq.~(\ref{NNLS1}) contains four terms, i.e., $G=G_{33}^s+G_{44}^s+G_{34}^d+G_{34}^e$; the nonlocal Kerr nonlinearities contributed by $G_{33}^s$, $G_{34}^d$, and $G_{34}^e$ are repulsive, but the one contributed by $G_{44}^s$ is attractive. Therefore, depending on system parameters and based on the competition among these four terms in $G$, the total Kerr nonlinearity of the system may be type of self-defocusing or self-focusing, which means that the system may support very rich nonlinear structures after the occurrence of the MI, including the emergence of various optical patterns and solitons.
Generally, when the repulsive part (contributed by $G_{33}^s$, $G_{34}^d$, and $G_{34}^e$)
plays a dominant role over the attractive part (contributed by $G_{44}^s$), the MI results in the formation of optical patterns; on the contrary, when the attractive part is dominant over the repulsive part, the MI gives rise to the formation of bright solitons.

As a first step, we focus on the case of pattern formation, for which the whole Kerr nonlinearity must be type of self-defocusing. This can be realized by choosing suitable system parameters to make the repulsive part in $G$\, (i.e., $G_{33}^s$, $G_{34}^d$, and $G_{34}^e$)  is larger than the attractive part (i.e., $G_{44}^s$). In fact, the system parameters given at the final part of Sec.~\ref{sec21} fulfill such requirement. Except for these parameters, other three parameters, i.e., $I_{\rm eff}$ (the effective probe-field intensity), $\sigma$ (the nonlocality degree of the Kerr nonlinearity), and $\Omega_m$ (the microwave field), play significant roles for determining the types of optical patterns in the system. Based on such consideration and for obtaining the optical patterns, we seek the ground-state solution of the system by a numerical simulation solving Eq.~(\ref{NNLS2}) via an imaginary evolution and split-step Fourier methods~\cite{YangJK2010}, for which the total energy of the system
{\small
\begin{align}
E=
&\int |\tilde{\nabla}_{\perp}u(\vec{\zeta},s)|^2 d^2\zeta\nonumber\\
&
+\frac{1}{2}\iint \Re({\vec{\zeta}^\prime-\vec{\zeta}})|u(\vec{\zeta},s)|^2
  |u(\vec{\zeta}^{\prime},s)|^2 d^2\zeta^{\prime}  d^2 \zeta\label{E}
\end{align}}\noindent
is minimum. The initial condition used in the simulation is the plane wave (\ref{PW}), perturbed by a random noise.

\begin{figure}
\centering
\includegraphics[width=0.48\textwidth]{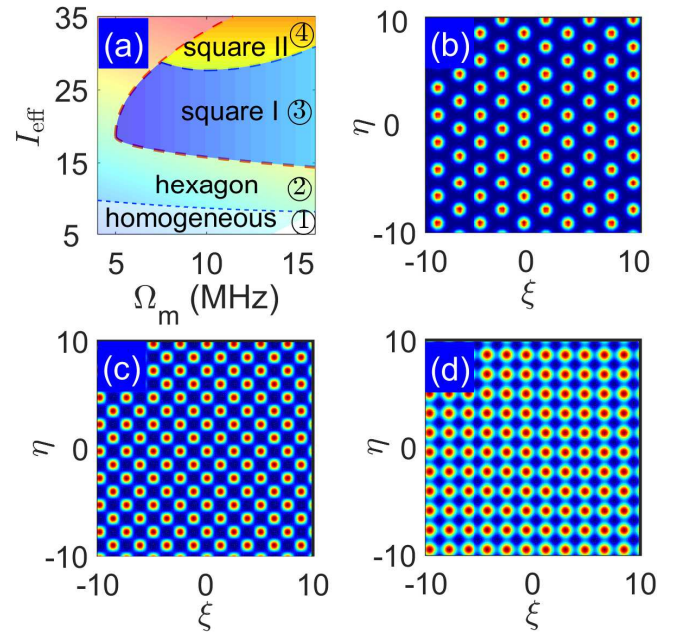}
\caption{\footnotesize Pattern formation and phase diagram controlled by the effective intensity of the probe field $I_{\rm eff}=\alpha A_0^2$
and the microwave field $\Omega_m$, for  the nonlocality degree $\sigma\equiv R_b/R_0=1$. Here, $A_0$ is the amplitude of the plane-wave state; $\alpha= -\int{\Re}(\vec{\zeta})d^2\zeta$; $R_b$ and $R_0$ are Rydberg blockade radius and the transverse radius of the probe beam, respectively.
(a)~Phase diagram of the structural transition of optical patterns, where different regions (phases) are obtained by changing the values of $I_{\rm eff}$ and $\Omega_m$.
Region $\circled{1}$: the homogeneous (i.e., the plane wave) state;
Region $\circled{2}$: the hexagonal lattice;
Region $\circled{3}$: the type I square lattice;
Region $\circled{4}$: the type II square lattice.
(b)~The hexagonal lattice of the normalized probe-field amplitude $|u|$ as a function of $\xi=x/R_0$ and $\eta=y/R_0$ for $\Omega_m=10$~MHz and $I_{\rm eff}=15$, corresponding to the region $\circled{2}$ in panel (a).
(c)~The type I square lattice of $|u|$ as a function of $\xi$ and $\eta$ for $\Omega_m=13$~MHz and $I_{\rm eff}=25$, corresponding to the region $\circled{3}$ in panel (a).
(d)~~The type II square lattice of $|u|$ as a function of $\xi$ and $\eta$ for $\Omega_m=15$~MHz and $I_{\rm eff}=35$, corresponding to the region $\circled{4}$ in panel (a).
}\label{Fig4}
\end{figure}

Shown in Fig.~\ref{Fig4}(a)
is the phase diagram describing the phase transition of self-organized optical structures, which are controlled by the effective intensity of the probe field $I_{\rm eff}=\alpha A_0^2$ and the microwave field $\Omega_m$.
The dashed lines in the figure are boundaries of different phases.
When obtaining the phase diagram, the nonlocality degree of the Kerr nonlinearity, i.e., $\sigma= R_b/R_0$, is fixed to be 1. From the figure, we see that several structural transitions of optical patterns
emerge when $I_{\rm eff}$ and $\Omega_m$ are changed in the following ways:
(i)~from the homogeneous state $\circled{1}$ to the hexagonal lattice $\circled{2}$;
(ii)~from the hexagonal lattice  $\circled{2}$ to the type I square lattice $\circled{3}$;
(iii)~from the type I square lattice $\circled{3}$ to the type II square lattice $\circled{4}$).
Here $\circled{1}$, $\circled{2}$, $\circled{3}$, and $\circled{4}$ represent regions of the homogeneous state, hexagonal lattice, type I square lattice, and type II square lattice, respectively.

To be more concrete, we give several examples for illustrating the optical lattice patterns that correspond to the self-organized structures indicated in the different regions of Fig.~\ref{Fig4}(a).  Fig.~\ref{Fig4}(b) shows a hexagonal lattice pattern, where the amplitude $|u|$ of the probe field is normalized  as a function of $\xi=x/R_0$ and $\eta=y/R_0$; it is obtained by taking $\Omega_m=10$~MHz and $I_{\rm eff}=15$, located in the region $\circled{2}$ of Fig.~\ref{Fig4}(a). Such hexagonal lattice pattern
was found by Sevinli {\it et al.}~\cite{Sevincli2011} where no microwave dressing is used (i.e., $\Omega_m=0$); in this case the hexagonal lattice pattern is the only one that can be obtained via the MI of the homogeneous (plane wave) state.

Plotted in Fig.~\ref{Fig4}(c) is the optical pattern by taking $|u|$ as a function of $\xi$ and $\eta$, for $\Omega_m=13$~MHz and $I_{\rm eff}=25$ [which locates in the region $\circled{3}$ of Fig.~\ref{Fig4}(a)]. We see that in this case a new optical structure, called the type I square lattice, emerges. Obviously, such new optical structure, which does not exist if the microwave dressing is absent, arises due to the symmetry breaking induced by the introduction of the microwave field.

Fig.~\ref{Fig4}(d) gives the result for the optical pattern with increasing microwave field and the effective intensity of the probe field,
by taking $\Omega_m=15$~MHz and $I_{\rm eff}=35$ [which is in the region
$\circled{4}$ of Fig.~\ref{Fig4}(a)]. One sees that in this situation another type of optical structure, called the type II square lattice, appears. By comparing the type I square lattice pattern of Fig.~\ref{Fig4}(c), we see that there is an angle difference (around $45^\circ$) between the type I and type II square lattices; furthermore, there are also differences for the normalized probe-field amplitudes and the lattice constants  between these two types of square lattice patterns (for detail, see Table~\ref{tab1} below).

\subsection{Pattern formation controlled by the nonlocality degree of the Kerr nonlinearity and the microwave field}\label{sec33}
To explore the structural phase transition of the optical patterns further, we now fix the effective probe-field intensity
($I_{\rm eff}=35$) but take the nonlocality degree of the Kerr nonlinearity $\sigma$ and the microwave field $\Omega_m$ as control parameters. Similar to the last subsection, we seek the spatial distribution of the probe field for which the total energy (\ref{E}) of the system is minimum, through a numerical simulation of Eq.~(\ref{NNLS2}).

\begin{figure}[htpb]
\centering
\includegraphics[width=0.49\textwidth]{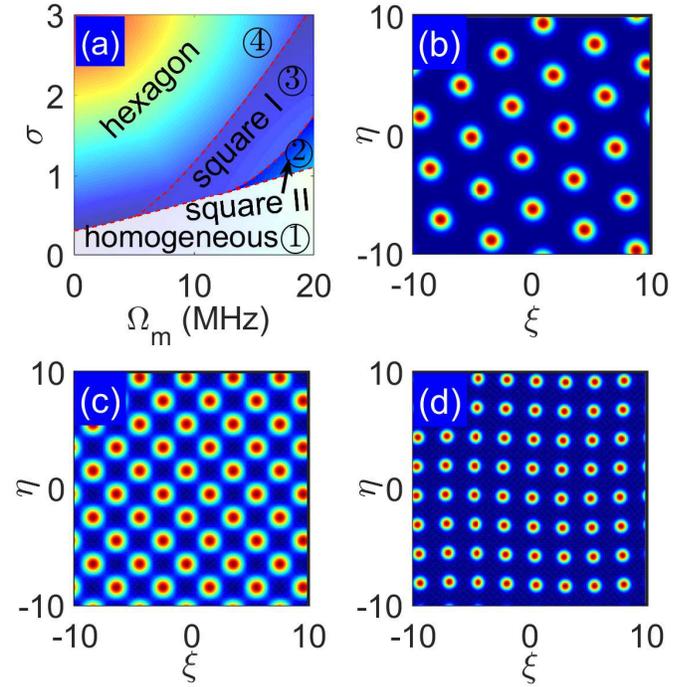}
\caption{\footnotesize Pattern formation and phase diagram controlled by the nonlocality degree of the Kerr nonlinearity $\sigma=R_b/R_0$ and the microwave field $\Omega_m$, for
the effective probe-field intensity $I_{\rm eff}=35$.
(a)~Phase diagram of the structural transition of optical patterns, where different regions (phases) are obtained by changing the values of $\sigma$ and $\Omega_m$.
Region $\circled{4}$: the pattern with hexagonal lattice structure;
Region $\circled{3}$: the pattern with type I square lattice structure;
Region $\circled{2}$: the pattern with type II square lattice structure;
Region $\circled{1}$: the homogeneous (i.e. the plane wave) state.
(b)~The hexagonal structure of the normalized probe-field amplitude $|u|$ as a function of $\xi=x/R_0$ and $\eta=y/R_0$ for $\Omega_m=10$~MHz and $\sigma=2$, corresponding to the region $\circled{4}$ in panel (a).
(c)~The type I square structure of $|u|$ as a function of $\xi$ and $\eta$ for $\Omega_m=12$~MHz and  $\sigma=1$, corresponding to the region $\circled{3}$ in panel (a).
(d)~~The type II square structure of $|u|$ as a function of $\xi$ and $\eta$ for $\Omega_m=18$~MHz and $\sigma=1$, corresponding to the region $\circled{4}$ in panel (a).
}\label{Fig5}
\end{figure}

Shown in Fig.~\ref{Fig5}(a) is the phase diagram of the structural transition of optical patterns, where different regions (phases) are obtained by changing the values of $\sigma$ and $\Omega_m$, separated by
dashed lines (i.e., boundaries of different phases).
We see that several structural transitions [i.e., from the homogeneous state $\circled{1}$ to the hexagonal lattice $\circled{4}$, from the hexagonal lattice $\circled{4}$ to the type I square lattice $\circled{3}$, and from the type I square lattice $\circled{3}$ to the type II square lattice $\circled{2}$\,] of the optical patterns arise  when $\sigma$ and $\Omega_m$ are varied.

We also give several examples for illustrating the optical patterns corresponding to the self-organized structures indicated in the different regions of Fig.~\ref{Fig5}(a). Fig.~\ref{Fig5}(b) shows a hexagonal lattice pattern, obtained by taking the normalized amplitude
of the probe field $|u|$ as a function of $\xi=x/R_0$ and $\eta=y/R_0$, for $\Omega_m=10$~MHz and $\sigma=2$ [located in the region $\circled{4}$ of Fig.~\ref{Fig5}(a)].

Shown in Fig.~\ref{Fig5}(c) is the optical pattern for $\Omega_m=12$~MHz and  $\sigma=1$ [located in the region $\circled{3}$ of Fig.~\ref{Fig5}(a)]; one sees that in this case the lattice pattern is a type I square lattice, which is absent without microwave field. Illustrated in Fig.~\ref{Fig5}(d) is the optical pattern with  $\Omega_m=18$~MHz and $\sigma=1$, which is in the region
$\circled{2}$ of Fig.~\ref{Fig5}(a); in this case
the type II square lattice structure appears.

To see clearly the differences between the two types of square lattice patterns, a quantitative comparison is made for the normalized probe-field amplitude $|u|$, microwave field $\Omega_m$, effective probe-field intensity $I_{\rm eff}$, and lattice constant $l$ (i.e., the distance between the maximums of two adjacent optical spots) between the type I and type II square lattice patterns obtained in Fig.~\ref{Fig4} and Fig.~\ref{Fig5} by taking the nonlocality degree of the Kerr nonlinearity $\sigma=1$, with the result presented in Table~\ref{tab1}.
\begin{table}[htbp]
\centering\caption{\footnotesize Differences between the type I and type II square lattice patterns. Values of the normalized probe-field amplitude $|u|$, microwave field $\Omega_m$, effective probe-field intensity $I_{\rm eff}$, and lattice constant $l$ for the two types of square lattice patterns.
}\label{tab1}
\vspace{4mm}
\renewcommand\tabcolsep{9.5pt}
\begin{tabular}{ccccc}
\Xhline{1pt}
Type
&$|u|_{\rm max}$
& $\Omega_m$
& $I_{\rm eff}$
& $l$\\
\Xhline{1pt}
Type I\,\,\,\, [Fig.~\ref{Fig4}(c)]
&28.5
&13
&25
&1.74\\
Type II\,\,\,\,[Fig.~\ref{Fig4}(d)]
&54.4
&15
&35
&1.66\\
Type I\,\,\,\, [Fig.~\ref{Fig5}(c)]
&37.9
&12
&35
&2.82\\
Type II\,\,\,\,[Fig.~\ref{Fig5}(d)]
&69.5
&18
&35
&2.35\\
\Xhline{1pt}
\end{tabular}%
\end{table}%
We see that:
(i)~the lattice constant $l$ of the type I square lattice pattern is larger than that of the type II one;
(ii)~comparing Fig.~\ref{Fig5}(c) with Fig.~\ref{Fig5}(d) [Fig.~\ref{Fig4}(c) with Fig.~\ref{Fig4}(d)], the lattice constant $l$
is larger for smaller microwave field $\Omega_m$.
The physical reason for such differences is that the nonlocal nonlinear response function $\Re/\Re_{\rm max}$ has a significant dependence on the microwave field. When the microwave filed $\Omega_m$ is increased, the shape of $\Re/\Re_{\rm max}$ is largely modified [i.e., it becomes narrower and steeper; see Fig.~\ref{Fig2}(d)], which makes the system change into a new state and thereby a new type of square lattice pattern emerges.

Combining Fig.~\ref{Fig4} and Fig.~\ref{Fig5}, which are the key results of this work, we see that, in the parameter domains considered here, the system supports three types of self-organized optical structures (i.e., the hexagonal lattice, the type I and type II square lattices), and their phase transitions can be controlled by actively manipulating the microwave field ($\Omega_m$), the effective probe-field intensity ($I_{\rm eff}$), and the nonlocality degree of the Kerr nonlinearity ($\sigma$).
The basic physical mechanism of the MI and the formation of the optical patterns found here may be understand as follows. When the plane-wave probe field with a finite amplitude is applied to and propagates in the Rydberg atomic gas along the $z$ direction, the nonlocal Kerr nonlinearity coming from the Rydberg-Rydberg interaction brings a phase modulation to the probe field; due to the role played by the diffraction in the transverse (i.e., $x$ and $y$) directions, the phase modulation is converted into  amplitude modulation. Because of the joint effect of the phase and amplitude modulations, in some parameter domains the probe field undergoes MI and re-organizes its spatial distribution and hence the formation of optical patterns occurs.

The emergence of the different self-organized structures (i.e. the hexagonal and square lattice optical patterns) and related phase changes are originated from the spatial symmetry breaking of the system.
To understand this and illustrate furthermore the differences between various optical lattice patterns, a detailed theoretical analysis on the ground-state energy of the system for different spatial distributions of the probe-field intensity is given in Appendix~\ref{appC}.

\subsection{Formation of nonlocal spatial optical solitons}\label{sec34}

Spatial optical solitons, i.e., localized nonlinear optical structures resulting from the balance between nonlinearity and diffraction, can form
through the MI of plane waves~\cite{Kivshar2006,Chen2012}. However,
a necessary condition for the formation of an optical soliton is that the Kerr nonlinearity in the system should be of the type of self-focusing. As indicated in Sec.~\ref{sec32}, due to the microwave dressing in our system there exist four kinds of nonlocal Kerr nonlinearities, which are described by the four response functions (i.e., $G_{33}^s$, $G_{44}^s$, $G_{34}^d$, $G_{34}^e$), and one of them (i.e., $G_{44}^s$) is attractive. Therefore, it is possible to make the total Kerr nonlinearity of the system to be a self-focused one if suitable system parameters are chosen.

In fact, a self-focused total Kerr nonlinearity can indeed be obtained by choosing the following system parameters, i.e.
$\Delta_2=6.28\times 10^2\,{\rm MHz}$,
$\Delta_3=6.92\,{\rm MHz}$,
${\rm \Delta_4=1\times 10^4\,Hz}$,
 ${\rm \Gamma_{12}=2\pi\times 6}$ MHz,
$\Gamma_3=\Gamma_{4}=2\pi\times 1.67\times 10^{-2}$ MHz, $\Omega_{c}=90\,{\rm MHz}$, and $\N_a=1.0\times 10^{11}\,{\rm
cm^{-3}}$.  In this situation, the attractive interaction contributed by $G_{44}^s$ plays a dominant role, and hence the plane-wave state (\ref{PW}) will undergo a MI and is possible to be squeezed into a soliton by the Kerr nonlinearity.

\begin{figure}
\centering
\includegraphics[width=0.45\textwidth]{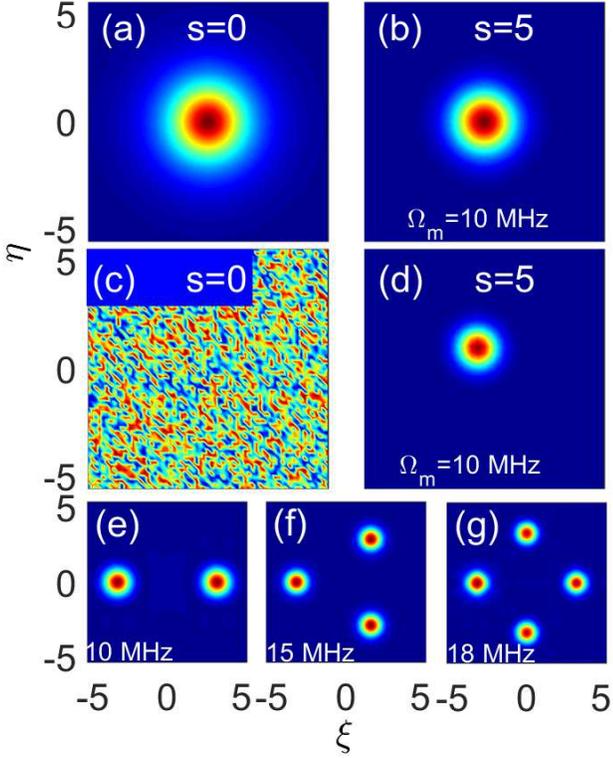}
\caption{\footnotesize Formation and propagation of nonlocal spatial optical solitons in the presence of the microwave dressing.
(a)~Initial condition ($|u|=1.3~\text{sech}[(\xi^2+\eta^2)^{1/2}]$ at $s=0$ ($s= z/2L_{\rm diff}$; $L_{\rm diff}$ is diffraction length).
(b)~Spatial distribution of the soliton when propagating to the position $s=5$, by taking the normalized probe-field amplitude $|u|$ as a function of non-dimensional coordinates $\xi= x/R_0$ and $\eta= y/R_0$
($R_0$ is the typical transverse beam radius of the probe field)
for the microwave field $\Omega_m=10$~MHz and the nonlocality degree $\sigma=1$.
(c)~Random initial condition $|u|= 1+ 0.05f$ ($f$ is a Gaussian noise)
and (d)~The soliton (generated by the random initial condition) when  propagating to $s=5$.
(e), (f), and (g) are two-, three-, and four-soliton solutions when they propagating to $s=5$ for $(\Omega_m, \sigma)=(10,1.2)$, $(15,1.4)$, and $(18,1.8)$, respectively.
}
\label{Fig6}
\end{figure}

To confirm the MI, a numerical simulation based on an imaginary-time propagation method is carried out by solving the NNLS equation (\ref{NNLS2}) with the above parameters. Shown in Fig.~\ref{Fig6}(b)
\noindent
is the spatial distribution of the probe-field envelope when it propagates 10 diffraction length (i.e., $s\equiv z/2L_{\rm diff}=5)$, by taking  $|u|$ (the normalized probe-field amplitude) as a function of non-dimensional coordinates $\xi= x/R_0$ and $\eta=y/R_0$, for the microwave field $\Omega_m=10$~MHz and the nonlocality degree of the Kerr nonlinearity $\sigma=1$. The initial condition used in the simulation is $|u|=1.3~\text{sech}[(\xi^2+\eta^2)^{1/2}]$\, [Fig.~\ref{Fig6}(a)]. We see that a nonlocal spatial optical soliton can indeed form in the system and it is quite stable during propagation.
Note that solitons can be also generated by using random initial conditions. Fig.~\ref{Fig6}(d) shows the spatial distribution of a soliton when it is created and propagates 10 diffraction length, for which the initial condition used is of the form $|u|= 1+ 0.05f$, where $f$ is Gaussian noise\, [Fig.~\ref{Fig6}(c)].

Different from the studies considered before, the nonlocal spatial optical solitons found here can be actively manipulated by actively tuning $\Omega_m$ and $\sigma$.  Shown in the panels (e), (f), and (g) of Fig.~\ref{Fig6} are two-soliton, three-soliton, and four-soliton solutions when they propagating to 10 diffraction length (i.e., $s=5$), obtained for $(\Omega_m, \sigma)=(10,1.2)$, $(15,1.4)$, and $(18,1.8)$, respectively.
Other multi-soliton solution solutions of the system may also be obtained.

\section{Summary}\label{sec4}

In this work, we have proposed a scheme for the realization of optical pattern formation and spatial solitons via a Rydberg-EIT. Through the use of a microwave dressing, we have shown that the nonlocal Kerr nonlinearity of the system can be manipulated actively and its magnitude can be enhanced significantly. Based on such nonlocal and tunable Kerr nonlinearity, we have demonstrated that a plane-wave probe field can undergo MI and spontaneous symmetry breaking, and thereby various self-organized optical patterns may emerge in the system. In particular, we have found that a hexagonal lattice pattern, which appears after the MI when the repulsive part of the nonlocal nonlinear response function is larger than its attractive part, may develop into several types of square lattice patterns when the microwave field is applied and tuned actively. Furthermore, through the MI the formation of nonlocal spatial optical solitons has also been found when the attractive part of the nonlocal nonlinear response function is dominant over its repulsive part. Different from the results reported before, the optical patterns and nonlocal optical solitons discovered here can be flexibly adjusted and controlled through the change of the effective probe-field intensity, nonlocality degree of the Kerr nonlinearity, and the strength of the microwave field. Our work opens a way for a versatile control of the self-organizations and structural phase transitions of laser light based on microwave-dressed Rydberg gases, which may have potential applications in optical information processing and transmission.

\acknowledgments

Authors thank Y.-C. Zhang and Z. Bai for useful discussions. This work was supported by the National Natural Science Foundation of China under Grant
No.~11474099 and No.~11975098.

\appendix
\begin{widetext}
\section{Bloch equations and solutions for density matrix elements}\label{app1}

\subsection{Explicit expressions of the Bloch equation for one-body density matrix elements}
The optical Bloch equation (\ref{Bloch0}) for the one-body density matrix (DM) elements $\rho_{\alpha\beta}({\bf r^{\prime}},t)=\langle\hat{S}_{\alpha\beta}({\bf r^{\prime}},t)\rangle$~\cite{note0} reads
{\small
\begin{subequations}
\begin{align}
&i\frac{\partial }{\partial t}\rho_{11}-i\Gamma_{12}\rho_{22}-\Omega_p\rho_{12}+\Omega_p^*\rho_{21}=0,\\
&i\left(\frac{\partial }{\partial t}+\Gamma_{12}\right)\rho_{22}-i\Gamma_{23}\rho_{33}
-i\Gamma_{24}\rho_{44}+\Omega_p\rho_{12}-\Omega_c\rho_{23}
-\Omega_p^*\rho_{21}+\Omega_c^*\rho_{32}=0,\\
&i\left(\frac{\partial }{\partial
t}+\Gamma_{23}\right)\rho_{33}+\Omega_c\rho_{23}
-\Omega_m\rho_{34}+\Omega_m^*\rho_{43}-\Omega_c^*\rho_{32}
+\mathcal{N}_a\int d^3{r}^\prime\mathcal{V}_{34}^e({\bf r}^\prime-{\bf r})\left[\rho_{43,34}({\bf r}^\prime,{\bf
r},t)-\rho_{34,43}({\bf r}^\prime,{\bf r},t)\right]=0,\\
&i\left(\frac{\partial }{\partial t}+\Gamma_{24}\right)\rho_{44}+\Omega_m\rho_{34}-\Omega_m^*\rho_{43}
-\mathcal{N}_a\int d^3{r}^\prime\mathcal{V}_{34}^e({\bf r}^\prime-{\bf r})\left[\rho_{43,34}({\bf r}^\prime, {\bf
r},t)-\rho_{34,43}({\bf r}^\prime,{\bf r},t)\right]=0,
\end{align}
\end{subequations}}
for diagonal matrix elements, and
{\small
\begin{subequations}
\begin{align}
&\left(i\frac{\partial }{\partial t}+d_{21}\right)\rho_{21}+\Omega_p(\rho_{11}-\rho_{22})+\Omega_c^*\rho_{31}=0,\\
&\left(i\frac{\partial }{\partial
t}+d_{31}\right)\rho_{31}+\Omega_c\rho_{21}+\Omega_m^*\rho_{41}
-\Omega_p\rho_{32}-\mathcal{N}_a\int d^3{
r}^\prime\mathcal{V}_{33}^s({\bf r}^\prime-{\bf r})\rho_{33,31}({\bf r}^\prime,{\bf  r},t)\notag\\
&\qquad-\mathcal{N}_a\int d^3{r}^\prime\mathcal{V}_{34}^d({ r}^\prime-{\bf r})\rho_{44,31}({\bf r}^\prime,{\bf
r},t)-\mathcal{N}_a\int d^3{r}^\prime\mathcal{V}_{34}^e({\bf r}^\prime-{\bf r})\rho_{34,41}({\bf r}^\prime, {\bf r},t)=0,\\
&\left(i\frac{\partial }{\partial
t}+d_{32}\right)\rho_{32}+\Omega_c(\rho_{22}-\rho_{33})
-\Omega_p^*\rho_{31}+\Omega_m^*\rho_{42}
-\mathcal{N}_a\int d^3{r}^\prime\mathcal{V}_{33}^s({\bf r}^\prime
-{\bf r})\rho_{33,32}({\bf r}^\prime,{\bf  r},t)\notag\\
&\qquad-\mathcal{N}_a\int d^3{r}^\prime\mathcal{V}_{34}^d({\bf r}^\prime-{\bf r})\rho_{44,32}({\bf r}^\prime,
{\bf r},t)-\mathcal{N}_a\int d^3{r}^\prime\mathcal{V}_{34}^e({\bf r}^\prime-{\bf r})\rho_{34,42}({\bf r}^\prime,{\bf r},t)=0,\\
&\left(i\frac{\partial }{\partial t}+d_{41}\right)\rho_{41}+\Omega_m\rho_{31}-\Omega_p\rho_{42}
-\mathcal{N}_a\int d^3{r'}\mathcal{V}_{44}^s({\bf r}^\prime-{\bf r})\rho_{44,41}({\bf r}^\prime,{\bf r},t)\notag\\
&\qquad-\mathcal{N}_a\int d^3{r}^\prime\mathcal{V}_{34}^d({\bf r}^\prime-{\bf r})\rho_{33,41}({\bf r}^\prime, {\bf
r},t)-\mathcal{N}_a\int d^3{r}^\prime\mathcal{V}_{34}^e({\bf r}^\prime-{\bf r})\rho_{43,31}({\bf r}^\prime,{\bf  r},t)=0,\\
&\left(i\frac{\partial }{\partial t}+d_{42}\right)\rho_{42}+\Omega_m\rho_{32}-\Omega_c\rho_{43}
-\Omega_p^*\rho_{41}
-\mathcal{N}_a\int d^3{r}^\prime\mathcal{V}_{44}^s({\bf r}^\prime-{\bf r})\rho_{44,42}({\bf r}^\prime,{\bf r},t),\notag\\
&\qquad-\mathcal{N}_a\int d^3{r'}\mathcal{V}_{34}^d({\bf r'-r})\rho_{33,42}({\bf r', r},t)-\mathcal{N}_a\int d^3{
r'}\mathcal{V}_{34}^e({\bf r'-r})\rho_{43,32}({\bf r', r},t)=0,\\
&\left(i\frac{\partial }{\partial t}+d_{43}\right)\rho_{43}+\Omega_m(\rho_{33}-\rho_{44})
-\Omega_c^*\rho_{42}
-\mathcal{N}_a\int d^3{r'}\left[\mathcal{V}_{44}^s({\bf r'-r})\rho_{44,43}({\bf  r',r},t)-\mathcal{V}_{33}^s({\bf
r'-r})\rho_{33,43}({\bf r},t)\right]\notag\\
&\qquad-\mathcal{N}_a\int d^3{r'}\mathcal{V}_{34}^d({\bf r'-r})\left[\rho_{33,43}({\bf r', r},t)-\rho_{44,43}({\bf
r',r},t)\right]
-\mathcal{N}_a\int d^3{r'}\mathcal{V}_{34}^e({\bf r'-r})
\left[\rho_{43,33}({\bf r',r},t)-\rho_{43,44}({\bf r',r},t)\right]=0,
\end{align}
\end{subequations}}
\end{widetext}
for non-diagonal matrix elements, where $d^3{r}'\equiv dx' dy' dz'$,  $\Gamma_3=\Gamma_{13}+\Gamma_{23}$, $d_{\alpha\beta}=\Delta_{\alpha}-\Delta_{\beta}+i \gamma_{\alpha\beta}~
(\alpha,\beta=1,2,3,4;\alpha \neq  \beta)$, with $\Delta_2=\omega_p-(\omega_2-\omega_1)$, $\Delta_3=
\omega_c+\omega_p-(\omega_3-\omega_1)$, and $\Delta_4=\omega_c+\omega_p+\omega_m-(\omega_4-\omega_1)$  the one-, two-, and three-photon detunings, respectively.
Here $\gamma_{\alpha\beta}=(\Gamma_{\alpha}+\Gamma_{\beta})/2
+\gamma_{\alpha\beta}^{\rm dep}$ with
$\Gamma_{\alpha}=\sum_{\alpha<\beta}\Gamma_{\alpha\beta}$, $\Gamma_{\alpha\beta}$ denoting the spontaneous emission decay rate from the state  $|\beta\rangle$ to the state
$|\alpha\rangle$ and $\gamma_{\alpha\beta}^{\rm dep}$ representing the dephasing rate reflecting the loss of phase coherence between
$|\alpha\rangle$ and $|\beta\rangle$.

From left hand side of the above equations, we see that, different from conventional EIT, there are many terms coming from the Rydberg-Rydberg interaction. One class of them involves the van der Waals interaction between the two atoms located respectively at positions ${\bf r}'$ and ${\bf r}$
and excited to the same Rydberg state [i.e., $\mathcal{V}_{\rm 33}^s({\bf r}'-{\bf r})$ and $\mathcal{V}_{44}^s({\bf r}'-{\bf r})$]; the other class involves the direct non-resonant van der Waals interaction and the
resonant exchange dipole-dipole interaction between the two atoms excited to different Rydberg states [i.e., $\mathcal{V}_{\rm 34}^{d}({\bf r}'-{\bf r})$ and $\mathcal{V}_{\rm 34}^e ({\bf r}'-{\bf r})]$.

Notice that, although the above equations describe the time evolution of the one-body DM elements $\rho_{\alpha\beta}({\bf r},t)$, they involve two-body DM elements $\rho_{\alpha\beta,\mu\nu}({\bf r^{\prime},r},t)=\langle
\hat{S}_{\alpha\beta}({\bf r^{\prime}},t) \hat{S}_{\mu\nu}({\bf r},t)\rangle$ due to the Rydberg-Rydberg interaction. Similarly, equations of motion for the two-body DM elements (not shown here for saving space) involves three-body DM elements, etc. To solve
such many-body problem, a suitable truncation for the infinite equation chain concerning many-body correlations is necessary, and a self-consistent calculation beyond mean-field approximation is needed, which have been developed recently~\cite{Bai2016,Zhang2018,Bai2019,Bai2020}. Based on such approach, we can solve the MB Eqs.~(\ref{Bloch0}) and
(\ref{Maxwell}) by using an asymptotic expansion in a standard way.

Notice also that in these equations there exists two types of nonlinearities. One of them (characterized by the terms like $\Omega ^*_{p}\rho_{31}$) arises from the photon-atom interaction due to the coupling between the probe field and atoms, which results in local Kerr nonlinearity if the two-photon detuning $\Delta_3$ is not zero~\cite{Bai2016,Zhang2018,Bai2019,Bai2020,Wang2001,Chen2014}; another one arises from the Rydberg-Rydberg interaction (characterized by the terms involving the two-body interactions potentials $\mathcal{V}_{33}^s$, $\mathcal{V}_{44}^s$, $\mathcal{V}_{\rm 34}^d$, and $\mathcal{V}_{\rm 34}^e$), which results in nonlocal Kerr nonlinearity. It is the nonlocal Rydberg-Rydberg interaction that makes the Rydberg-EIT interesting and typical on the study of nonlocal nonlinear optics.

\subsection{Solutions for density matrix elements}

We are interested in stationary states of the
system, and hence the time derivatives in the MB equations (\ref{Bloch0}) and (\ref{Maxwell}) can be neglected
(i.e.,~$\partial/\partial t= 0$), which is valid if the probe, control, and microwave fields have large time durations. We adopt the method developed in Refs.~\cite{Bai2016,Zhang2018,Bai2019,Bai2020} to firstly solve the Bloch equation (\ref{Bloch0}) under the condition of Rydberg-EIT. We assume that all the atoms are initially prepared in the ground state $|1\rangle$.

Since the probe field is assumed to be weak, one can take $\Omega_{p}\, (\sim\epsilon$) as an expansion parameter, and
$\rho_{\alpha\alpha}=\delta_{\alpha,1}\rho_{\alpha\alpha}^{(0)}+\epsilon \rho_{\alpha\alpha}^{\left(1\right)}+\epsilon
^2\rho_{\alpha\alpha}^{\left(2\right)}+\cdots$, $\rho_{\alpha\beta}=\epsilon \rho_{\alpha\beta}^{\left(1
\right)}+\epsilon ^2\rho_{\alpha\beta}^{\left( 2 \right)}+\cdots$, $\left(\beta=1,2,3;\alpha=1,2,3,4;\beta<\alpha \right)$. Substituting these expansions into Eq.~(\ref{Bloch0}) and collecting coefficients of
the same power of $\epsilon$, we can solve equations for $\rho_{\alpha\beta}^{(m)}$ ($m=1,2,3,...$) order by order.

\vspace{5mm}
\emph{(1). First-order solution.} The first-order equation reads
\begin{align}\label{a21}
  \left(
     \begin{array}{cccc}
       d_{21} & \Omega_c^* & 0 \\
       \Omega_c & d_{31} &\Omega_m^*\\
        0 & \Omega_m&d_{41} \\
     \end{array}
   \right)
   \left(\begin{array}{c}
            \rho_{21}^{(1)}\\
           \rho_{31}^{(1)} \\
           \rho_{41}^{(1)}
         \end{array}
   \right)
   =   \left(\begin{array}{c}
            -\Omega_p\\
          0\\
          0
         \end{array}
   \right).
\end{align}
Its solution is given by
$\rho_{21}^{(1)}=-(D_m/D)\Omega_p\equiv  a_{21}^{(1)}\Omega_p$,
$\rho_{31}^{(1)}=-(d_{41}\Omega_c/{D})\Omega_p\equiv  a_{31}^{(1)}\Omega_p$, and $\rho_{41}^{(1)}=(\Omega_m\Omega_c/{D})\Omega_p\equiv  a_{41}^{(1)}\Omega_p$ with $D=d_{21}D_m+d_{41}|\Omega_c|^2$, and
\begin{equation}\label{Dm}
D_m=|\Omega_m|^2-d_{31}d_{41}.
\end{equation}

\vspace{5mm}
\emph{(2). Second-order solution.} At this order, the solution of nonzero DM elements is given by
$\rho_{\alpha\alpha}^{(2)}=a_{\alpha\alpha}^{(2)}|\Omega_p|^2$, ($\alpha=1,2,3,4$), and
$\rho_{\alpha\beta}^{(2)}=a_{\alpha\beta}^{(2)}|\Omega_p|^2$~($\alpha=4,3$; $\beta=3,2$), where the coefficients $a_{\alpha\alpha}^{(2)}$ and $a_{\alpha\beta}^{(2)}$ can be obtained by solving the
equation
{\small
\begin{align}
  &\left(
  \begin{array}{ccccccccc}
    0&i\Gamma_{23}& 0&\Omega_m^*&  -\Omega_m& 0& 0& -\Omega_c^*& \Omega_c\\
    0& 0& i\Gamma_{24}& -\Omega_m^*& \Omega_m& 0& 0& 0& 0\\
    0& -\Omega_c& 0& 0& 0& \Omega_m^*& 0& d_{32}& 0\\
    0& 0 &0 &-\Omega_c& 0& d_{42}& 0& \Omega_m& 0\\
    0& \Omega_m& -\Omega_m& d_{43} &0 &-\Omega_c^* &0 &0 &0\\
    0& -\Omega_c& 0& 0& 0& 0& \Omega_m& 0& d_{32}^*\\
    0& 0& 0& 0& -\Omega_c^*& 0& d_{42}^*& 0& \Omega_m^*\\
    0& \Omega_m^*& -\Omega_m^* &0 &d_{43}^*& 0 &-\Omega_c& 0& 0\\
    1 &1& 1& 0& 0& 0& 0 &0& 0
  \end{array}
  \right)\notag\\
  &\times\left(\begin{array}{c}
  \rho_{11}^{(2)}\\
  \rho_{33}^{(2)}\\
  \rho_{44}^{(2)}\\
  \rho_{43}^{(2)}\\
  \rho_{34}^{(2)}\\
  \rho_{42}^{(2)}\\
  \rho_{24}^{(2)}\\
  \rho_{32}^{(2)}\\
  \rho_{23}^{(2)}
\end{array}
\right)
=\left(
  \begin{array}{c}
    0 \\
    0 \\
    \Omega_p^{*}\rho_{31}^{(1)}-\Omega_c\rho_{22}^{(2)} \\
    \Omega_p^{*}\rho_{41}^{(1)} \\
    0 \\
    \Omega_p\rho_{31}^{*(1)}-\Omega_c^*\rho_{22}^{(2)}  \\
    \Omega_p\rho_{41}^{*(1)} \\
    0\\
    \rho_{22}^{(2)}
  \end{array}
  \right),
\end{align}
with $\rho_{22}^{(2)}=[\Omega_p^{*(1)}\rho_{21}^{(1)}
-\Omega_p^{(1)}\rho_{21}^{*(1)}]/(i\Gamma_{12})\equiv a_{22}^{(2)}|\Omega_p|^2$.

\vspace{5mm}
\emph{(3). Third-order solution.}
Third order solution for $\rho_{21}^{(3)}$ reads
{\small\begin{align}
\rho_{21}^{(3)}&=\frac{\mathcal{A}}{D}
|\Omega_p|^2\Omega_p -\frac{\Omega_c\Omega_m\mathcal{N}_a}{D}\int d^3{r'} \left[ \mathcal{V}_{44}^s\rho_{44,41}^{(3)}({\bf r}^\prime,{\bf
r},t)\right.\notag\\&\left.+\mathcal{V}_{34}^d\rho_{33,41}^{(3)}({\bf r}^\prime,{\bf r},t)+\mathcal{V}_{34}^e\rho_{43,31}^{(3)}({\bf r}^\prime,{\bf
r},t)\right]\notag\\
&+\frac{\Omega_cd_{41}\mathcal{N}_a}{D}\int d^3{r'}
\left[\mathcal{V}_{33}^s\rho_{33,31}^{(3)}({\bf r',r},t)\right.\notag\\&\left.+\mathcal{V}_{34}^d\rho_{44,31}^{(3)}({\bf
r',r},t)+\mathcal{V}_{34}^e\rho_{34,41}^{(3)}({\bf r',r},t)\right],
\end{align}}\noindent
with
\begin{equation}\label{A}
\mathcal{A}=D_m(a_{22}^{(2)}-a_{11}^{(2)})
+d_{41}\Omega_ca_{32}^{(2)}-\Omega_c\Omega_ma_{42}^{(2)}.
\end{equation}
Results for $\rho_{31}^{(3)}$ and $\rho_{41}^{(3)}$ have similar forms, but omitted here for saving space.

Notice that for obtaining the solutions of $\rho_{21}^{(3)}$, $\rho_{31}^{(3)}$, and $\rho_{41}^{(3)}$, equations for some two-body DM elements $\rho_{\alpha\beta,\mu\nu}$ must be solved simultaneously. These two-body DM elements
are nonzero starting at $\epsilon^2$-order, so they can be assumed to have the form $\rho_{\alpha\beta,\mu\nu}=\epsilon^2 \rho^{(2)}_{\alpha\beta,\mu\nu}+\epsilon^3 \rho^{(3)}_{\alpha\beta,\mu\nu}+\cdots$. Then we have the equation
{\small\begin{align}
  \left(\begin{matrix}
    M_1 & \Omega_c& \Omega_m^* &0& 0  \\
    \Omega_c^*&M_2 & 0 & \Omega_m^* &0 \\
    \Omega_m &0 & M_3 & \Omega_c  & \Omega_m^* \\
    0 & \Omega_m & \Omega_c^*  & M_4 & 0  \\
   0 & 0 & \Omega_m &  0 & M_5
\end{matrix}
  \right)
  \begin{pmatrix}
    \rho_{31,31}^{(2)}\\
    \rho_{31,21}^{(2)}\\
    \rho_{41,31}^{(2)}\\
    \rho_{41,21}^{(2)}\\
    \rho_{41,41}^{(2)}\\
    \end{pmatrix}
  =\left(\begin{matrix}
  0\\
  -\Omega_p\rho_{31}^{(1)}\\
  0\\
  -\Omega_p\rho_{41}^{(1)}\\
  0\\
  \end{matrix}
  \right),
\end{align}}\noindent
where $M_1=d_{31}-\mathcal{V}_{33}^s/2$, $M_2=d_{21}+d_{31}$, $M_3=d_{41}+d_{31}-\mathcal{V}_{34}^d
-\mathcal{V}_{34}^e$, $M_4=d_{41}+d_{21}$, and $M_5=d_{41}-\mathcal{V}_{44}^s/2$. The solution reads $\rho_{\alpha
1,\beta1}^{(2)}=a_{\alpha 1,\beta1}^{(2)}|\Omega_p|^2$~($\alpha,\beta=2,3,4$), where $a_{\alpha 1,\beta1}^{(2)}$ are constants (their explicit expressions are omitted here).

With these results, the third-order equations of the two-body DM elements (which are too lengthy thus omitted here) can be solved, which have the solution of the form $\rho_{\alpha \beta,\mu\nu}^{(3)}=a_{\alpha
\beta,\mu\nu}^{(3)}|\Omega_p({\bf r}',t)|^2\Omega_p({\bf r},t)$, where $a_{\alpha \beta,\mu\nu}^{(3)}$ are functions of ${\bf r}'-{\bf r}$.
Notice that, when solving the equations of the two-body DM elements, some three-body DM elements are involved.
To make the equations closed, the method developed in Ref.~\cite{Bai2016,Zhang2018,Bai2019,Bai2020} has been exploited to factorize the three-body DM elements into one- and two-body ones.

Solutions of the equations of density matrix elements above the third-order approximation can also be obtained in a similar way.
However, we can stop here since in this work we are interested only in the Kerr effect up to third order. Thereby, based on the first-, second-, and third-order solutions given above, we may obtain the explicit expression of $\rho_{21}=\rho_{21}^{(1)}+\rho_{21}^{(2)}+\rho_{21}^{(3)}$ (by setting $\epsilon=1$).

\section{Expression of the third-order nonlinear optical susceptibility and the derivation of the nonlocal NLS equation}\label{app3}
The optical susceptibility of the probe field is defined by
$\chi=\mathcal{N}_a({\bf e\cdot p}_{12})\rho_{21}/(\epsilon_0\mathcal{E}_p)$. Based on the result in the Appendix \ref{app1}, we have $\chi=\chi^{(1)}+\left[\chi^{(3)}_{\rm loc}+\chi^{(3)}_{\rm nloc}\right]|\mathcal{E}_p|^2$~\cite{note100}, with the linear and local third-order nonlinear susceptibilities given by
{\small
\begin{eqnarray}\label{chi3s}
&& \chi^{(1)}= \frac{\mathcal{N}_a|{\bf p}_{12}|^2D_m}{\epsilon_0\hbar D}, \label{chi3s1}\\
&& \chi_{\rm loc}^{(3)}=\frac{\mathcal{N}_a|{\bf p}_{12}|^4\mathcal{A}}{\epsilon_0\hbar^3 D},\label{chi3s2}
\end{eqnarray}
}\noindent
respectively, where $D_m$ and $\mathcal{A}$ are given by (\ref{Dm}) and (\ref{A}). Due to the Rydberg-Rydberg interaction, the system also supports the  nonlocal third-order nonlinear susceptibility
\begin{align}\label{chi3n}
\chi_{\rm nloc}^{(3)}=&\frac{\mathcal{N}^2_a|{\bf p}_{12}|^4}{\epsilon_0\hbar^3}\frac{\Omega_c}{D}\int d^3{
r'} \notag\\
&\times\bigg\{d_{41}\big[a_{33,31}^{(3)}\mathcal{V}_{33}^s+a_{44,31}^{(3)}
\mathcal{V}_{34}^d+a_{34,41}^{(3)}\mathcal{V}_{34}^e\big]\notag\\
&-\Omega_m\big[a_{44,41}^{(3)}\mathcal{V}_{44}^s
+a_{33,41}^{(3)}\mathcal{V}_{34}^d+a_{43,31}^{(3)}
\mathcal{V}_{34}^e\big]\bigg\}.
\end{align}
where $\mathcal{V}_{\alpha\beta}^l\equiv \mathcal{V}_{\alpha\beta}^l({\bf r'})$ $(l=s,d,e)$ and $a_{\alpha\beta,\mu\nu}^{(3)}\equiv a_{\alpha\beta,\mu\nu}^{(3)}({\bf r}',t)$.

Substituting the result of $\rho_{21}$ obtained in the Appendix~\ref{app1} into the Maxwell equation (\ref{Maxwell}), we obtain the envelope
equation for $\Omega_p$, which has the form of
the nonlocal nonlinear Schr\"odinger (NNLS) equation
\begin{align}\label{W1}
&i\frac{\partial \Omega_p }{\partial z} +\frac{c}{2\omega_p}\nabla_{\perp}^{2}\Omega_p + {W_1}|\Omega_p{|^2}\Omega_p\notag\\
&\hspace{1.5cm}+\int d^3{
r'} R({\bf r'-{\bf r}})|\Omega_p({\bf r'})|^2\Omega_p({\bf r})=0,
\end{align}\noindent
where
$W_1=\mathcal{A}\kappa_{12}/D$ and
$R({\bf r})=\sum_{\alpha=3,4}R_{\alpha\alpha}^s({\bf r})+\sum_{\gamma=d,e} R_{34}^\gamma({\bf r})$ with
\begin{align}\label{eqb5}
R_{33}^s ({\bf r})&=\kappa_{12}\Omega_cd_{41}\mathcal{N}_a a_{33,31}^{(3)}({\bf r})\mathcal{V}_{33}^s({\bf r})/D,\notag\\
R_{44}^s({\bf r})&=-\kappa_{12}\Omega_c\Omega_m\mathcal{N}_a a_{44,41}^{(3)}({\bf r})\mathcal{V}_{44}^s({\bf r})/D,\notag\\
R_{34}^d({\bf r})&=\kappa_{12}\Omega_c\mathcal{N}_a [\Omega_m
a_{33,41}^{(3)}({\bf r})-d_{41}a_{44,31}^{(3)}({\bf r})]\mathcal{V}_{34}^d({\bf r})/D, \notag\\
R_{34}^e({\bf r})&=-\kappa_{12}\Omega_c\mathcal{N}_a [\Omega_m
a_{43,31}^{(3)}({\bf r})-d_{41}a_{34,41}^{(3)}({\bf r})]\mathcal{V}_{34}^e({\bf r})/D,
\end{align}
which are contributed by
$\mathcal{V}_{33}^s$, $\mathcal{V}_{44}^s$, $\mathcal{V}_{\rm 34}^d$, and $\mathcal{V}_{\rm 34}^e$, respectively.
The coefficient $W_1$ characterizes the local self-phase modulation and the nonlinear response function $R({\bf r})$ characterizes nonlocal self-phase modulation of the probe field.

For simplicity, we assume that the probe field is slowly varied along
the $z$ direction, so that a local approximation in this direction can be made for the nonlinear response function. Then Eq.~(\ref{W1}) is reduced into
{\small \begin{align}
&i\frac{\partial \Omega_p }{\partial z} +\frac{c}{2\omega_p}\nabla_{\perp}^{2}\Omega_p + {W_1}|\Omega_p{|^2}\Omega_p\notag\\
& \hspace{0cm}+\int d^2{
r'}G({\bf r_{\perp}'-r_{\perp}}) |\Omega_p({\bf
r_{\perp}'},z)|^2\Omega_p({\bf r_{\perp}},z)=0,\label{NNLSE1}
\end{align}}\noindent
where $ {\bf r}_{\perp}=(x,y)$ and $d^2
r'=dx'dy'$, the reduced nonlocal nonlinear response function reads
${G}({\bf r_{\perp}})=G_{33}^s({\bf r_{\perp}})+G_{44}^s({\bf r_{\perp}})+G_{34}^d({\bf  r_{\perp}})+G_{34}^e({\bf r_{\perp}})$, with $G_{\alpha\beta}^l({\bf r_{\perp}})=\int R_{\alpha\beta}^l({\bf r})
dz$ ($\{\alpha\beta\}=\{ 33,44,34\}$;  $l=s,d,e$).

Equation (\ref{NNLSE1}) can be written into the dimensionless form
{\small\begin{align}\label{NLS}
& i\frac{\partial u}{\partial s} + \tilde{\nabla}_{\perp}^2u +w_1|u{|^2}u\nonumber\\
&+\int d^2\zeta' \Re(\vec{\zeta}'-\vec{\zeta})|u(\vec{\zeta}',s){|^2}u(\vec{\zeta},s)=0,
\end{align}}\noindent
where $u=\Omega_p/U_0$  ($U_0$ is the typical half Rabi frequency of the probe field), $s = z/(2{L_{\rm diff}})$  (${L_{\rm diff}}={\omega _p}R_0^2/c$ is the typical diffraction length), $w_1= 2U_0^2L_{\rm diff}R_0^2W_1$, $\Re(\vec{\zeta}'-\vec{\zeta})=2L_{\rm diff}U_0^2R_0^2 G[{ (\vec{\zeta}'-\vec{\zeta})}R_0]$,  $\tilde{\nabla}_{\perp}^2=\partial^2/\partial\xi^2+\partial^2/\partial\eta^2$, $\vec{\zeta}=(\xi,\eta)= (x,y)/{R_0}$ ($R_0$  the typical transverse radius of the probe beam), and $d^2\zeta'= d\xi' d\eta'$.

\section{Ground-state energy analysis on the hexagonal and square lattice patterns}\label{appC}

Here we make a detailed analysis to illustrate the reason why the system supports only the hexagonal, type I, and type II square lattices, found in the parameter range we used. To this end, we assume that the solution of the NNLS Eq.~(\ref{NNLS2}) is a superposition of many Fourier modes, i.e.,
\begin{align}\label{FourierS}
u(\vec{\zeta})=\sum_{j=1}^N A_j\,e^{i\vec{\beta}_j\cdot \vec{\zeta}},
\end{align}
where $\vec{\beta}_j=(\beta_{1j},\beta_{2j})$ are non-dimensional wavenumbers, and $A_j$ ($j=1,2,...,N$) are complex amplitudes
which are functions of $s$, $\xi$, and $\eta$ when the Kerr nonlinearity
of the system plays a significant role.

Because the system has a rotation symmetry, generally the value of $N$ can take a very large value. However, due to the joint action of the Kerr nonlinearity and diffraction, the system undergoes MI for some wavenumbers
and hence symmetry breaking, by which only several modes are kept at end. To see this, we assume all the modes in (\ref{FourierS}) are unstable ones, and satisfy $|\vec{\beta}_j|\approx \beta_{\rm cr}$ where $\beta_{\rm cr}$ is the first wavenumber of the unstable band of the MI [e.g., in the shadow regions of Fig.~\ref{Fig3}(a)].

\begin{figure}[htpb]
\centering
\includegraphics[width=0.45\textwidth]{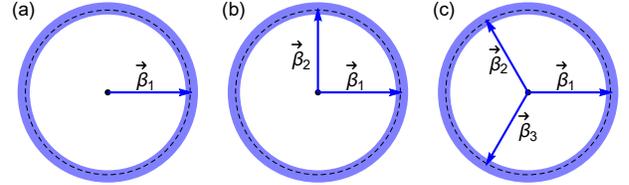}
\caption{\footnotesize  Schematic diagram of the non-dimensional wavevectors ${\vec \beta}_j$ for different optical self-organized structures. (a)~Parallel stripes; (b)~Square lattice, for which ${\vec \beta}_1\cdot {\vec \beta}_2=0$; (c)~Hexagonal lattice, for which ${\vec \beta}_1+{\vec \beta}_2+{\vec \beta}_3=0$. All the wavevectors have the same module (i.e.
$|\vec{\beta}_j|\approx \beta_{\rm cr}$; $\beta_{\rm cr}$ is the first wavenumber of the unstable band of the MI).
}
\label{FigS1}
\end{figure}

Any complicated periodic pattern are made of the most periodic basic pattern, i.e., an array of parallel stripes (or called rolls). For example, a square lattice pattern is made of two kinds of parallel stripes with equal amplitudes but different orientations (the angle difference between the directions of the two kinds of parallel stripes is $90^\circ$), a hexagonal pattern is made of three kinds of parallel stripes with equal amplitudes but different orientations (the angle difference between any two kinds of parallel stripes is $120^\circ$).

Shown in panels (a), (b) and (c) of Fig.~\ref{FigS1}
are non-dimensional wavevectors for the parallel stripes, square lattice (for which ${\vec \beta}_1\cdot {\vec \beta}_2=0$), and  hexagonal lattice (for which ${\vec \beta}_1+{\vec \beta}_2+{\vec \beta}_3=0$). Note that here all the wavevectors have the same module (i.e., $|\vec{\beta}_j|\approx \beta_{\rm cr}$; $\beta_{\rm cr}$ is the first wavenumber of the unstable band of the MI).

In order to determine which lattice pattern arises first under given system parameters, we consider the total energy of the system
{\small
\begin{align}
E=
&\int |\tilde{\nabla}_{\perp}u(\vec{\zeta},s)|^2 d^2\zeta\nonumber\\
&
+\frac{1}{2}\iint \Re({\vec{\zeta}^\prime-\vec{\zeta}})|u(\vec{\zeta},s)|^2
  |u(\vec{\zeta}^{\prime},s)|^2 d^2\zeta^{\prime}  d^2 \zeta.\label{s1}
\end{align}}\noindent
Note that the response function $\Re$ has an imaginary part, which however  is very small due to the EIT effect and hence is negligible
in the calculation below.

For hexagonal lattice pattern, we assume the solution of the NNLS equation (\ref{NNLS2}) has the form
\begin{eqnarray}\label{s2}
&& u=\sqrt{\rho}e^{i\phi}e^{i\mu s}, \label{s21}\\
&& \rho=\rho_0\big[1+\sum_{j=1}^3(D_j\,e^{i\vec{\beta}_j
   \cdot\vec{\zeta}}+{\rm c.c.})\big],\label{s22}
\end{eqnarray}
where $\rho_0=A_0^2$;
$\mu=-\int \Re(\vec{\zeta})d^2\zeta$;
$D_j$ ($j=1,2,3$) are complex constants called modulation amplitudes;
the wavevectors $\vec{\beta}_j$ fulfill the condition
$\vec{\beta}_1+\vec{\beta}_2+\vec{\beta}_3=0$ with
$|\beta_1|=|\beta_2|=|\beta_3|=\beta_{\rm cr}$. Note that for obtaining
the solution with minimal energy, the phase $\phi$  must be homogeneous (i.e. a real constant)~\cite{Pomeau1994,Pomeau2007}.

\begin{figure}[htpb]
\centering
\includegraphics[width=0.45\textwidth]{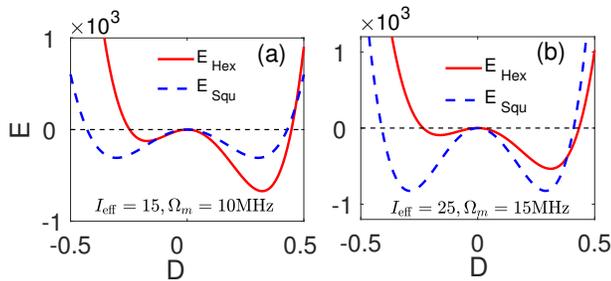}
\caption{
\footnotesize Ground-state energy for cases of hexagonal and square lattices.
(a)~Ground-state energy  $E$ as a function of modulation amplitude $D$, with parameters $\beta_{\rm cr}=4.3, I_{\rm eff}=15$,  $\Omega_m=10$~MHz, and $\sigma=1$. The solid red line ($E_{\rm Hex}$) is for the hexagonal pattern; the dotted blue line ($E_{\rm Squ}$) is for the square pattern.
(b)~The same as (a) but for $\beta_{\rm cr}=4.3, I_{\rm eff}=25$, $\Omega_m=15$~MHz, and $\sigma=1$.
}
\label{S2}
\end{figure}
\noindent

Inserting (\ref{s21}) and (\ref{s22}) into Eq.~(\ref{s1}) and expanding $D_j$ in Taylor series, we obtain the energy of the system for the hexagonal lattice
\begin{align}\label{s3}
  E_{\text{Hex}}=&\rho_0 \beta_{\rm cr}^2 V\Big[
  \frac{1}{2}\sum_{j=1}^3(|D_j|^2+|D_j|^4)-\frac{3}{4}(D_1D_2D_3
  +D_1^*D_2^*D_3^*)\notag\\
  &+2(|D_1|^2|D_2|^2+|D_1|^2|D_3|^2+|D_2|^2|D_3|^2)
  \Big]\notag\\
  & +\frac{1}{2}\rho_0^2V\tilde{\Re}({\beta_{\rm cr}})\sum_{j=1}^3|D_j|^2+\text{high order term},
\end{align}
where $\tilde{\Re}({\beta_{\rm cr}})$ is the value of the response function $\Re$ in momentum space for $\beta=\beta_{\rm cr}$, $V= \int d^2{\zeta}$ is the volume of the system.  The
ground-state energy can be acquired by solving the equations $\partial E_{\text{Hex}}/\partial D_j=0$ ($j=1,2,3)$. Then, we obtain its expression
\begin{align}
   {E_{\text{Hex}}}=&\rho_0{V} \beta_{\rm cr}^2 \Big[
  \frac{3}{2}(D^2-D^3)+\frac{15}{2}D^4\Big]
  +\frac{3}{2}V\rho_0^2\tilde{\Re}({\beta_{\rm cr}})D^2,
\end{align}
for $D_{1}=D_{2}=D_{3}=D$.
The ground-state energy for square lattice pattern can be obtained in a similarly way, which reads
\begin{align}
  {E_{\text{Squ}}}={V}\rho_0 \beta_{\rm cr}^2 (D^2+3D^4) +V\rho_0^2\tilde{\Re}({\beta_{\rm cr}})D^2.
\end{align}

Shown in Fig.~\ref{S2}(a) is the ground-state energy $E$ as a function of modulation amplitude $D$, with parameters $\beta_{\rm cr}=4.3, I_{\rm eff}=15$, $\Omega_m=10$~MHz, and $\sigma=1$. The solid red line in the figure is for the ground-state energy $E_{\rm Hex}$ of the hexagonal lattice; the dotted blue line is the ground-state energy $E_{\rm Squ}$ for the square lattice. We see that the minimal energy of the system occurs at $D=0.32$ (where $E_{\rm min}=E_{\rm Hex,min}=-0.78\times10^3$), which means that the hexagonal lattice pattern is preferred to emerge in the system. The case shown in Fig.~\ref{S2}(b) is similar to Fig.~\ref{S2}(a), but the parameters are taken as $\beta_{\rm cr}=4.3, I_{\rm eff}=25$, and $\Omega_m=15$~MHz. We see that in this case
the minimal energy of the system occurs for $D=\pm 0.28$ (where $E_{\rm min}=E_{\rm Squ,min}=-0.81\times10^3$), which means that the system favors the  emergence of the square lattice pattern.
These results are consistent with the ones found in Fig.~\ref{Fig4}
and Fig.~\ref{Fig5}.



\begin{references}
\bibitem{note0}
We assume all the atoms are initially populate in the state $|1\rangle$, and hence the average of the operator $\hat{O}$ amounts to $\langle \hat{O}\rangle=\langle G|\hat{O}|G\rangle$, with $|G\rangle=|1,1,1,...,1\rangle$.
\bibitem{note100}
In the calculation of the Kerr coefficients given here, we have assumed that the spatial variation of the probe-field envelope ${\cal E}_p$ is slow and thus ${\cal E}_p$ can be approximately taken to be a constant.
\end{references}
\end{document}